\documentclass[dvipsnames,format=sigconf]{acmart}

\setcopyright{acmcopyright}
\copyrightyear{2018}
\acmYear{2018}
\acmDOI{10.1145/1122445.1122456}

\copyrightyear{2023} 
\acmYear{2023} 
\setcopyright{acmlicensed}\acmConference[GECCO '23]{Genetic and Evolutionary Computation Conference}{July 15--19, 2023}{Lisbon, Portugal}
\acmBooktitle{Genetic and Evolutionary Computation Conference (GECCO '23), July 15--19, 2023, Lisbon, Portugal}
\acmPrice{15.00}
\acmDOI{10.1145/3583131.3590448}
\acmISBN{979-8-4007-0119-1/23/07}

\usepackage{xspace}
\usepackage[capitalise]{cleveref}
\usepackage{graphicx}
\usepackage{subcaption}
\usepackage[%
  group-minimum-digits=4,%
  list-final-separator={, and },%
  add-integer-zero=false,%
  free-standing-units,%
  round-mode=figures,%
  round-precision=3,%
  binary-units,%
  detect-weight=true,%
  detect-inline-weight=math,%
]{siunitx}

\newcommand{\SpeedOverBaselineProbThirty}{169}
\newcommand{\SpeedOverBaselineProbSixty}{72}
\newcommand{\SpeedOverBaselineProbHundred}{52}

\newcommand{\MeanCovRandomCatchTheDots}{95.81}

\newcommand{\MeanCovProbZeroCatchTheDots}{95.81}

\newcommand{\MeanCovProbThirtyCatchTheDots}{96.14}

\newcommand{\MeanCovProbSixtyCatchTheDots}{95.85}
\newcommand{\MeanCovProbHundredCatchTheDots}{96.18}

\newcommand{\MeanCovRandomFinalFight}{97.44}

\newcommand{\MeanCovProbZeroFinalFight}{100}

\newcommand{\MeanCovProbThirtyFinalFight}{100}

\newcommand{\MeanCovProbSixtyFinalFight}{100}
\newcommand{\MeanCovProbHundredFinalFight}{100}

\newcommand{\MeanCovRandomFlappyParrot}{92.63}

\newcommand{\MeanCovProbZeroFlappyParrot}{95.53}

\newcommand{\MeanCovProbThirtyFlappyParrot}{96.58}

\newcommand{\MeanCovProbSixtyFlappyParrot}{94.82}
\newcommand{\MeanCovProbHundredFlappyParrot}{94.56}

\newcommand{\MeanCovRandomFruitCatching}{69.39}

\newcommand{\MeanCovProbZeroFruitCatching}{99.14}

\newcommand{\MeanCovProbThirtyFruitCatching}{100}

\newcommand{\MeanCovProbSixtyFruitCatching}{100}
\newcommand{\MeanCovProbHundredFruitCatching}{100}

\newcommand{\MeanCovRandomHackAttack}{92.47}

\newcommand{\MeanCovProbZeroHackAttack}{98.43}

\newcommand{\MeanCovProbThirtyHackAttack}{99.25}

\newcommand{\MeanCovProbSixtyHackAttack}{99.25}
\newcommand{\MeanCovProbHundredHackAttack}{\textbf{99.50}}

\newcommand{\MeanCovRandomPong}{91.11}

\newcommand{\MeanCovProbZeroPong}{99.89}

\newcommand{\MeanCovProbThirtyPong}{100}

\newcommand{\MeanCovProbSixtyPong}{100}
\newcommand{\MeanCovProbHundredPong}{100}

\newcommand{\MeanCovRandomSnake}{95.61}

\newcommand{\MeanCovProbZeroSnake}{94.79}

\newcommand{\MeanCovProbThirtySnake}{94.44}

\newcommand{\MeanCovProbSixtySnake}{94.94}
\newcommand{\MeanCovProbHundredSnake}{\textbf{95.22}}

\newcommand{\MeanCovRandomSpaceOdyssey}{93.10}

\newcommand{\MeanCovProbZeroSpaceOdyssey}{95.28}

\newcommand{\MeanCovProbThirtySpaceOdyssey}{97.67}

\newcommand{\MeanCovProbSixtySpaceOdyssey}{\textbf{99.05}}
\newcommand{\MeanCovProbHundredSpaceOdyssey}{\textbf{100}}
\newcommand{\TotalMeanCovRandom}{90.95}
\newcommand{\TotalMeanCovProbZero}{97.36}
\newcommand{\TotalMeanCovProbThirty}{98.01}
\newcommand{\TotalMeanCovProbSixty}{97.99}
\newcommand{\TotalMeanCovProbHundred}{98.18}

\newcommand{\MeanCovWaitInfCatchTheDots}{97.76}

\newcommand{\MeanCovWaitHundredCatchTheDots}{94.80}
\newcommand{\MeanCovWaitTenCatchTheDots}{94.35}

\newcommand{\MeanCovWaitInfFinalFight}{100}

\newcommand{\MeanCovWaitHundredFinalFight}{100}
\newcommand{\MeanCovWaitTenFinalFight}{100}

\newcommand{\MeanCovWaitInfFlappyParrot}{93.51}

\newcommand{\MeanCovWaitHundredFlappyParrot}{93.95}
\newcommand{\MeanCovWaitTenFlappyParrot}{94.25}

\newcommand{\MeanCovWaitInfFruitCatching}{100}

\newcommand{\MeanCovWaitHundredFruitCatching}{100}
\newcommand{\MeanCovWaitTenFruitCatching}{100}

\newcommand{\MeanCovWaitInfHackAttack}{94.73}

\newcommand{\MeanCovWaitHundredHackAttack}{95.73}
\newcommand{\MeanCovWaitTenHackAttack}{95.86}

\newcommand{\MeanCovWaitInfPong}{100}

\newcommand{\MeanCovWaitHundredPong}{100}
\newcommand{\MeanCovWaitTenPong}{100}

\newcommand{\MeanCovWaitInfSnake}{95.39}

\newcommand{\MeanCovWaitHundredSnake}{94.39}
\newcommand{\MeanCovWaitTenSnake}{93.47}

\newcommand{\MeanCovWaitInfSpaceOdyssey}{99.54}

\newcommand{\MeanCovWaitHundredSpaceOdyssey}{100}
\newcommand{\MeanCovWaitTenSpaceOdyssey}{96.31}
\newcommand{\TotalMeanCovWaitInf}{97.62}
\newcommand{\TotalMeanCovWaitHundred}{97.36}
\newcommand{\TotalMeanCovWaitTen}{96.78}

\newcommand{\MeanCovThirtySecCatchTheDots}{96.67}

\newcommand{\MeanCovOneMinCatchTheDots}{96.18}

\newcommand{\MeanCovThirtySecFinalFight}{100}

\newcommand{\MeanCovOneMinFinalFight}{100}

\newcommand{\MeanCovThirtySecFlappyParrot}{94.91}

\newcommand{\MeanCovOneMinFlappyParrot}{94.56}

\newcommand{\MeanCovThirtySecFruitCatching}{100}

\newcommand{\MeanCovOneMinFruitCatching}{100}

\newcommand{\MeanCovThirtySecHackAttack}{98.75}

\newcommand{\MeanCovOneMinHackAttack}{99.50}

\newcommand{\MeanCovThirtySecPong}{100}

\newcommand{\MeanCovOneMinPong}{100}

\newcommand{\MeanCovThirtySecSnake}{95.33}

\newcommand{\MeanCovOneMinSnake}{95.22}

\newcommand{\MeanCovThirtySecSpaceOdyssey}{99.77}

\newcommand{\MeanCovOneMinSpaceOdyssey}{100}

\newcommand{\TotalMeanCovThirtySec}{98.18}
\newcommand{\TotalMeanCovOneMin}{98.18}

\newcommand{\SamplesOneMinCatchTheDots}{30\xspace}

\newcommand{\WaitPropOneMinCatchTheDots}{0.00\xspace}
\newcommand{\SamplesOneMinFinalFight}{149\xspace}

\newcommand{\WaitPropOneMinFinalFight}{0.00\xspace}
\newcommand{\SamplesOneMinFlappyParrot}{99\xspace}

\newcommand{\WaitPropOneMinFlappyParrot}{0.00\xspace}
\newcommand{\SamplesOneMinFruitCatching}{66\xspace}

\newcommand{\WaitPropOneMinFruitCatching}{0.00\xspace}
\newcommand{\SamplesOneMinHackAttack}{166\xspace}

\newcommand{\WaitPropOneMinHackAttack}{0.00\xspace}
\newcommand{\SamplesOneMinPong}{72\xspace}

\newcommand{\WaitPropOneMinPong}{0.00\xspace}
\newcommand{\SamplesOneMinSnake}{38\xspace}

\newcommand{\WaitPropOneMinSnake}{0.00\xspace}
\newcommand{\SamplesOneMinSpaceOdyssey}{59\xspace}

\newcommand{\WaitPropOneMinSpaceOdyssey}{0.00\xspace}
\newcommand{\SamplesThirtySecCatchTheDots}{13\xspace}

\newcommand{\WaitPropThirtySecCatchTheDots}{0.00\xspace}
\newcommand{\SamplesThirtySecFinalFight}{88\xspace}

\newcommand{\WaitPropThirtySecFinalFight}{0.00\xspace}
\newcommand{\SamplesThirtySecFlappyParrot}{52\xspace}

\newcommand{\WaitPropThirtySecFlappyParrot}{0.00\xspace}
\newcommand{\SamplesThirtySecFruitCatching}{33\xspace}

\newcommand{\WaitPropThirtySecFruitCatching}{0.00\xspace}
\newcommand{\SamplesThirtySecHackAttack}{69\xspace}

\newcommand{\WaitPropThirtySecHackAttack}{0.00\xspace}
\newcommand{\SamplesThirtySecPong}{25\xspace}

\newcommand{\WaitPropThirtySecPong}{0.00\xspace}
\newcommand{\SamplesThirtySecSnake}{22\xspace}

\newcommand{\WaitPropThirtySecSnake}{0.00\xspace}
\newcommand{\SamplesThirtySecSpaceOdyssey}{36\xspace}

\newcommand{\WaitPropThirtySecSpaceOdyssey}{0.00\xspace}
\newcommand{\SamplesWaitTenCatchTheDots}{206\xspace}

\newcommand{\WaitPropWaitTenCatchTheDots}{0.53\xspace}
\newcommand{\SamplesWaitTenFinalFight}{498\xspace}

\newcommand{\WaitPropWaitTenFinalFight}{0.30\xspace}
\newcommand{\SamplesWaitTenFlappyParrot}{562\xspace}

\newcommand{\WaitPropWaitTenFlappyParrot}{0.49\xspace}
\newcommand{\SamplesWaitTenFruitCatching}{310\xspace}

\newcommand{\WaitPropWaitTenFruitCatching}{0.49\xspace}
\newcommand{\SamplesWaitTenHackAttack}{436\xspace}

\newcommand{\WaitPropWaitTenHackAttack}{0.09\xspace}
\newcommand{\SamplesWaitTenPong}{242\xspace}

\newcommand{\WaitPropWaitTenPong}{0.37\xspace}
\newcommand{\SamplesWaitTenSnake}{272\xspace}

\newcommand{\WaitPropWaitTenSnake}{0.53\xspace}
\newcommand{\SamplesWaitTenSpaceOdyssey}{342\xspace}

\newcommand{\WaitPropWaitTenSpaceOdyssey}{0.35\xspace}
\newcommand{\SamplesWaitHundredCatchTheDots}{132\xspace}

\newcommand{\WaitPropWaitHundredCatchTheDots}{0.27\xspace}
\newcommand{\SamplesWaitHundredFinalFight}{428\xspace}

\newcommand{\WaitPropWaitHundredFinalFight}{0.03\xspace}
\newcommand{\SamplesWaitHundredFlappyParrot}{239\xspace}

\newcommand{\WaitPropWaitHundredFlappyParrot}{0.00\xspace}
\newcommand{\SamplesWaitHundredFruitCatching}{178\xspace}

\newcommand{\WaitPropWaitHundredFruitCatching}{0.05\xspace}
\newcommand{\SamplesWaitHundredHackAttack}{369\xspace}

\newcommand{\WaitPropWaitHundredHackAttack}{0.01\xspace}
\newcommand{\SamplesWaitHundredPong}{143\xspace}

\newcommand{\WaitPropWaitHundredPong}{0.00\xspace}
\newcommand{\SamplesWaitHundredSnake}{167\xspace}

\newcommand{\WaitPropWaitHundredSnake}{0.23\xspace}
\newcommand{\SamplesWaitHundredSpaceOdyssey}{264\xspace}

\newcommand{\WaitPropWaitHundredSpaceOdyssey}{0.03\xspace}
\newcommand{\SamplesWaitNoneThreeMinCatchTheDots}{109\xspace}

\newcommand{\WaitPropWaitNoneThreeMinCatchTheDots}{0.00\xspace}
\newcommand{\SamplesWaitNoneThreeMinFinalFight}{566\xspace}

\newcommand{\WaitPropWaitNoneThreeMinFinalFight}{0.00\xspace}
\newcommand{\SamplesWaitNoneThreeMinFlappyParrot}{276\xspace}

\newcommand{\WaitPropWaitNoneThreeMinFlappyParrot}{0.00\xspace}
\newcommand{\SamplesWaitNoneThreeMinFruitCatching}{189\xspace}

\newcommand{\WaitPropWaitNoneThreeMinFruitCatching}{0.00\xspace}
\newcommand{\SamplesWaitNoneThreeMinHackAttack}{370\xspace}

\newcommand{\WaitPropWaitNoneThreeMinHackAttack}{0.00\xspace}
\newcommand{\SamplesWaitNoneThreeMinPong}{107\xspace}

\newcommand{\WaitPropWaitNoneThreeMinPong}{0.00\xspace}
\newcommand{\SamplesWaitNoneThreeMinSnake}{113\xspace}

\newcommand{\WaitPropWaitNoneThreeMinSnake}{0.00\xspace}
\newcommand{\SamplesWaitNoneThreeMinSpaceOdyssey}{202\xspace}

\newcommand{\WaitPropWaitNoneThreeMinSpaceOdyssey}{0.00\xspace}
\newcommand{\WinningStatesRandomCatchTheDots}{5\xspace}
\newcommand{\WinningStatesRandomFinalFight}{0\xspace}
\newcommand{\WinningStatesRandomFlappyParrot}{0\xspace}
\newcommand{\WinningStatesRandomFruitCatching}{0\xspace}
\newcommand{\WinningStatesRandomHackAttack}{0\xspace}
\newcommand{\WinningStatesRandomPong}{10\xspace}
\newcommand{\WinningStatesRandomSnake}{0\xspace}
\newcommand{\WinningStatesRandomSpaceOdyssey}{0\xspace}
\newcommand{\WinningStatesAverageRandom}{1.88\xspace}
\newcommand{\WinningStatesZeroProbCatchTheDots}{13\xspace}
\newcommand{\WinningStatesZeroProbFinalFight}{30\xspace}
\newcommand{\WinningStatesZeroProbFlappyParrot}{10\xspace}
\newcommand{\WinningStatesZeroProbFruitCatching}{20\xspace}
\newcommand{\WinningStatesZeroProbHackAttack}{13\xspace}
\newcommand{\WinningStatesZeroProbPong}{29\xspace}
\newcommand{\WinningStatesZeroProbSnake}{0\xspace}
\newcommand{\WinningStatesZeroProbSpaceOdyssey}{1\xspace}
\newcommand{\WinningStatesAverageZeroProb}{14.50\xspace}
\newcommand{\WinningStatesThirtyProbCatchTheDots}{13\xspace}
\newcommand{\WinningStatesThirtyProbFinalFight}{30\xspace}
\newcommand{\WinningStatesThirtyProbFlappyParrot}{13\xspace}
\newcommand{\WinningStatesThirtyProbFruitCatching}{30\xspace}
\newcommand{\WinningStatesThirtyProbHackAttack}{27\xspace}
\newcommand{\WinningStatesThirtyProbPong}{30\xspace}
\newcommand{\WinningStatesThirtyProbSnake}{0\xspace}
\newcommand{\WinningStatesThirtyProbSpaceOdyssey}{19\xspace}
\newcommand{\WinningStatesAverageThirtyProb}{20.25\xspace}
\newcommand{\WinningStatesSixtyProbCatchTheDots}{16\xspace}
\newcommand{\WinningStatesSixtyProbFinalFight}{30\xspace}
\newcommand{\WinningStatesSixtyProbFlappyParrot}{7\xspace}
\newcommand{\WinningStatesSixtyProbFruitCatching}{30\xspace}
\newcommand{\WinningStatesSixtyProbHackAttack}{27\xspace}
\newcommand{\WinningStatesSixtyProbPong}{30\xspace}
\newcommand{\WinningStatesSixtyProbSnake}{0\xspace}
\newcommand{\WinningStatesSixtyProbSpaceOdyssey}{25\xspace}
\newcommand{\WinningStatesAverageSixtyProb}{20.62\xspace}
\newcommand{\WinningStatesHundredProbCatchTheDots}{14\xspace}
\newcommand{\WinningStatesHundredProbFinalFight}{30\xspace}
\newcommand{\WinningStatesHundredProbFlappyParrot}{4\xspace}
\newcommand{\WinningStatesHundredProbFruitCatching}{30\xspace}
\newcommand{\WinningStatesHundredProbHackAttack}{28\xspace}
\newcommand{\WinningStatesHundredProbPong}{30\xspace}
\newcommand{\WinningStatesHundredProbSnake}{5\xspace}
\newcommand{\WinningStatesHundredProbSpaceOdyssey}{30\xspace}
\newcommand{\WinningStatesAverageHundredProb}{21.38\xspace}
\newcommand{\WinningStatesSizeThirtysecCatchTheDots}{14\xspace}
\newcommand{\WinningStatesSizeThirtysecFinalFight}{30\xspace}
\newcommand{\WinningStatesSizeThirtysecFlappyParrot}{4\xspace}
\newcommand{\WinningStatesSizeThirtysecFruitCatching}{30\xspace}
\newcommand{\WinningStatesSizeThirtysecHackAttack}{25\xspace}
\newcommand{\WinningStatesSizeThirtysecPong}{30\xspace}
\newcommand{\WinningStatesSizeThirtysecSnake}{4\xspace}
\newcommand{\WinningStatesSizeThirtysecSpaceOdyssey}{28\xspace}
\newcommand{\WinningStatesAverageSizeThirtysec}{20.62\xspace}
\newcommand{\WinningStatesSizeOneminCatchTheDots}{14\xspace}
\newcommand{\WinningStatesSizeOneminFinalFight}{30\xspace}
\newcommand{\WinningStatesSizeOneminFlappyParrot}{4\xspace}
\newcommand{\WinningStatesSizeOneminFruitCatching}{30\xspace}
\newcommand{\WinningStatesSizeOneminHackAttack}{28\xspace}
\newcommand{\WinningStatesSizeOneminPong}{30\xspace}
\newcommand{\WinningStatesSizeOneminSnake}{5\xspace}
\newcommand{\WinningStatesSizeOneminSpaceOdyssey}{30\xspace}
\newcommand{\WinningStatesAverageSizeOnemin}{21.38\xspace}

\newcommand{\WinningStatesWaitTenCatchTheDots}{7\xspace}
\newcommand{\WinningStatesWaitTenFinalFight}{30\xspace}
\newcommand{\WinningStatesWaitTenFlappyParrot}{6\xspace}
\newcommand{\WinningStatesWaitTenFruitCatching}{30\xspace}
\newcommand{\WinningStatesWaitTenHackAttack}{14\xspace}
\newcommand{\WinningStatesWaitTenPong}{30\xspace}
\newcommand{\WinningStatesWaitTenSnake}{0\xspace}
\newcommand{\WinningStatesWaitTenSpaceOdyssey}{3\xspace}
\newcommand{\WinningStatesAverageWaitTen}{15.00\xspace}
\newcommand{\WinningStatesWaitHundredCatchTheDots}{9\xspace}
\newcommand{\WinningStatesWaitHundredFinalFight}{30\xspace}
\newcommand{\WinningStatesWaitHundredFlappyParrot}{1\xspace}
\newcommand{\WinningStatesWaitHundredFruitCatching}{30\xspace}
\newcommand{\WinningStatesWaitHundredHackAttack}{13\xspace}
\newcommand{\WinningStatesWaitHundredPong}{30\xspace}
\newcommand{\WinningStatesWaitHundredSnake}{1\xspace}
\newcommand{\WinningStatesWaitHundredSpaceOdyssey}{30\xspace}
\newcommand{\WinningStatesAverageWaitHundred}{18.00\xspace}
\newcommand{\WinningStatesWaitInfCatchTheDots}{17\xspace}
\newcommand{\WinningStatesWaitInfFinalFight}{30\xspace}
\newcommand{\WinningStatesWaitInfFlappyParrot}{1\xspace}
\newcommand{\WinningStatesWaitInfFruitCatching}{30\xspace}
\newcommand{\WinningStatesWaitInfHackAttack}{6\xspace}
\newcommand{\WinningStatesWaitInfPong}{30\xspace}
\newcommand{\WinningStatesWaitInfSnake}{4\xspace}
\newcommand{\WinningStatesWaitInfSpaceOdyssey}{27\xspace}
\newcommand{\WinningStatesAverageWaitInf}{18.12\xspace}

\newcommand\definetool[2]{\newcommand{#1}{{\textsc{#2}}\xspace}}
\definetool{\Scratch}{Scratch}
\definetool{\Whisker}{Whisker}
\definetool{\WhiskerWeb}{WhiskerWeb}
\definetool{\Neatest}{Neatest}

\usepackage[
    colorinlistoftodos,prependcaption,textsize=tiny
]{todonotes}

\begin{document}

\title{Learning by Viewing: Generating Test Inputs for Games by Adapting Neural Networks to Human Gameplay Traces}

\title{Learning by Viewing: Generating Test Inputs for Games by Integrating Human Gameplay Traces in Neuroevolution}

% Authors
\author{Patric Feldmeier}
\affiliation{%
  \institution{University of Passau}
  \country{Germany}}
\email{patric.feldmeier@uni-passau.de}

\author{Gordon Fraser}
\affiliation{%
  \institution{University of Passau}
  \country{Germany}}
\email{gordon.fraser@uni-passau.de}

%%
%% The abstract is a short summary of the work to be presented in the
%% article.
\begin{abstract}
% Context
Although automated test generation is common in many programming domains, games still challenge test generators due to their heavy randomisation and hard-to-reach program states.
% Problem
Neuroevolution combined with search-based software testing principles has been shown to be a promising approach for testing games, but the co-evolutionary search for optimal network topologies and weights involves unreasonably long search durations.
% Contribution
In this paper, we aim to improve the evolutionary search for game input generators by integrating knowledge about human gameplay behaviour.
To this end, we propose a novel way of systematically recording human gameplay traces, and integrating these traces into the evolutionary search for networks using traditional gradient descent as a mutation operator.
% Results
Experiments conducted on eight diverse \Scratch games demonstrate that the proposed approach reduces the required search time from five hours down to only \SpeedOverBaselineProbHundred\ minutes.

\end{abstract}

%% The code below is generated by the tool at http://dl.acm.org/ccs.cfm.


%% Keywords. 

\maketitle
\section{Introduction}
\begin{figure}[!tbp]
  \begin{subfigure}[b]{.49\columnwidth}
    \centering
    \includegraphics[width=\columnwidth]{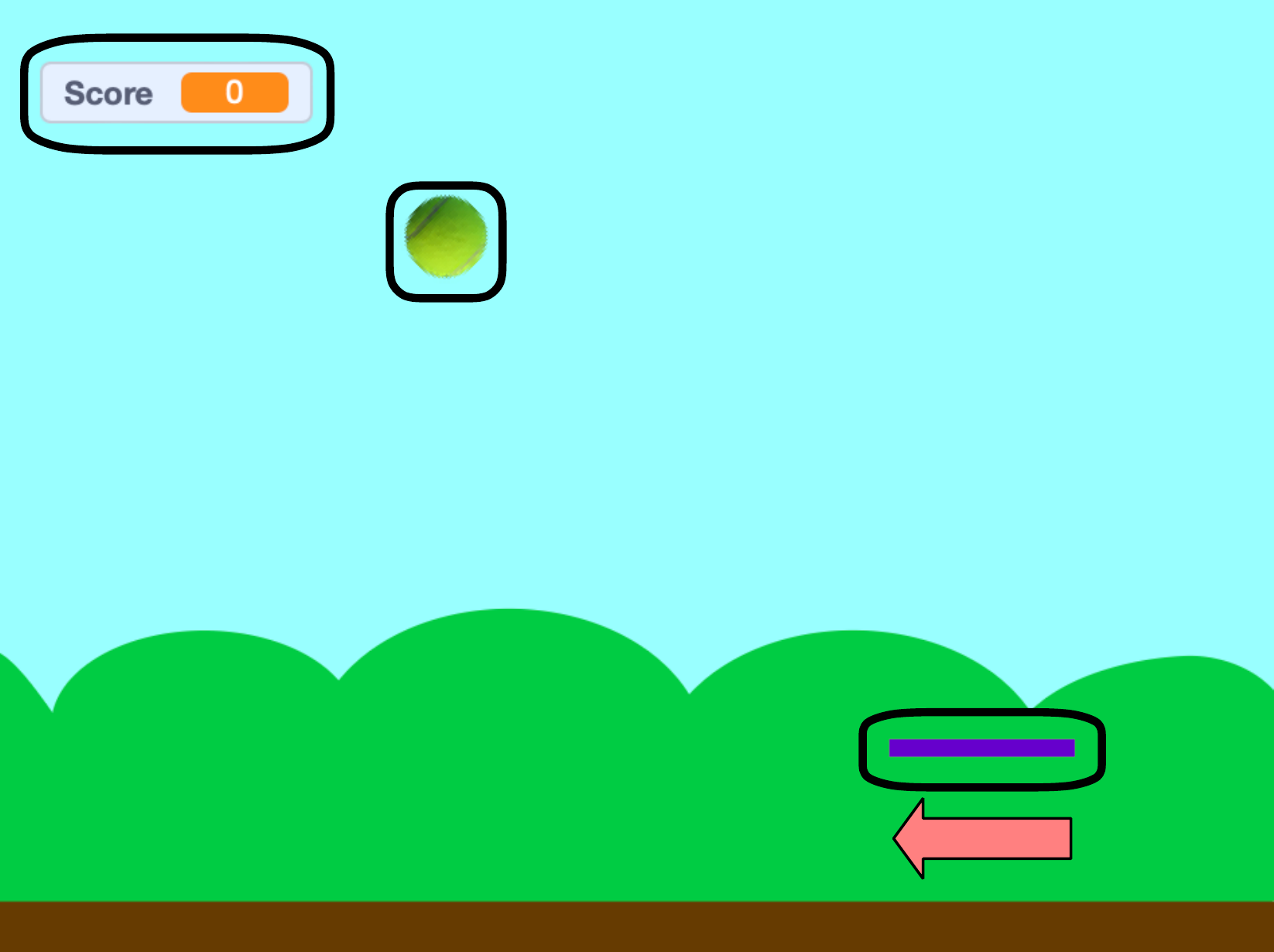}
    \caption{Player moves to the left}
    \label{fig:Pong-1}
  \end{subfigure}%
  \hspace{.1em}
  \begin{subfigure}[b]{.49\columnwidth}
      \centering
    \includegraphics[width=\columnwidth]{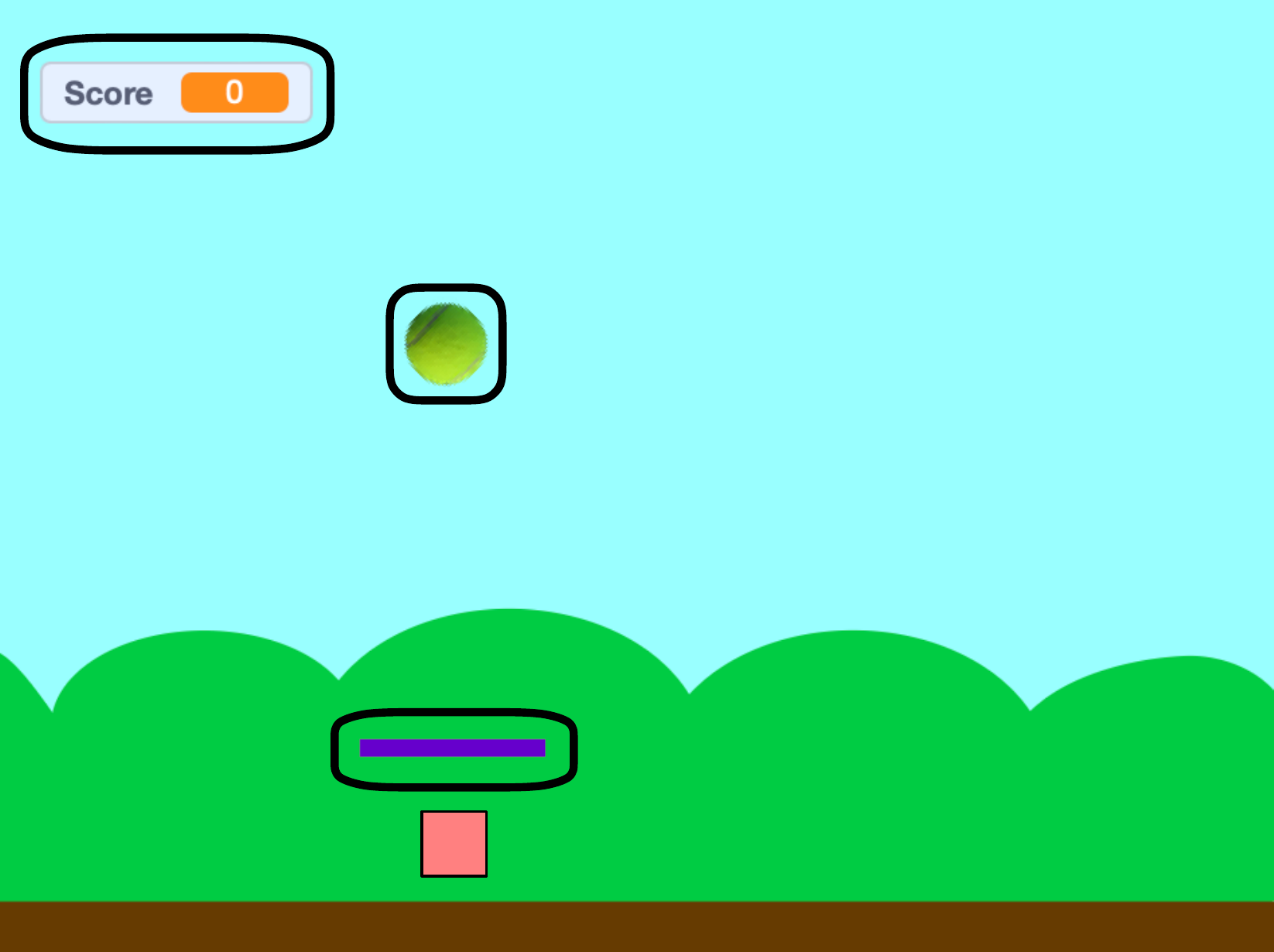}
    \caption{Player performs no action}
    \label{fig:Pong-2}
  \end{subfigure}
  \begin{subfigure}[b]{.49\columnwidth}
    \centering
    \includegraphics[width=\columnwidth]{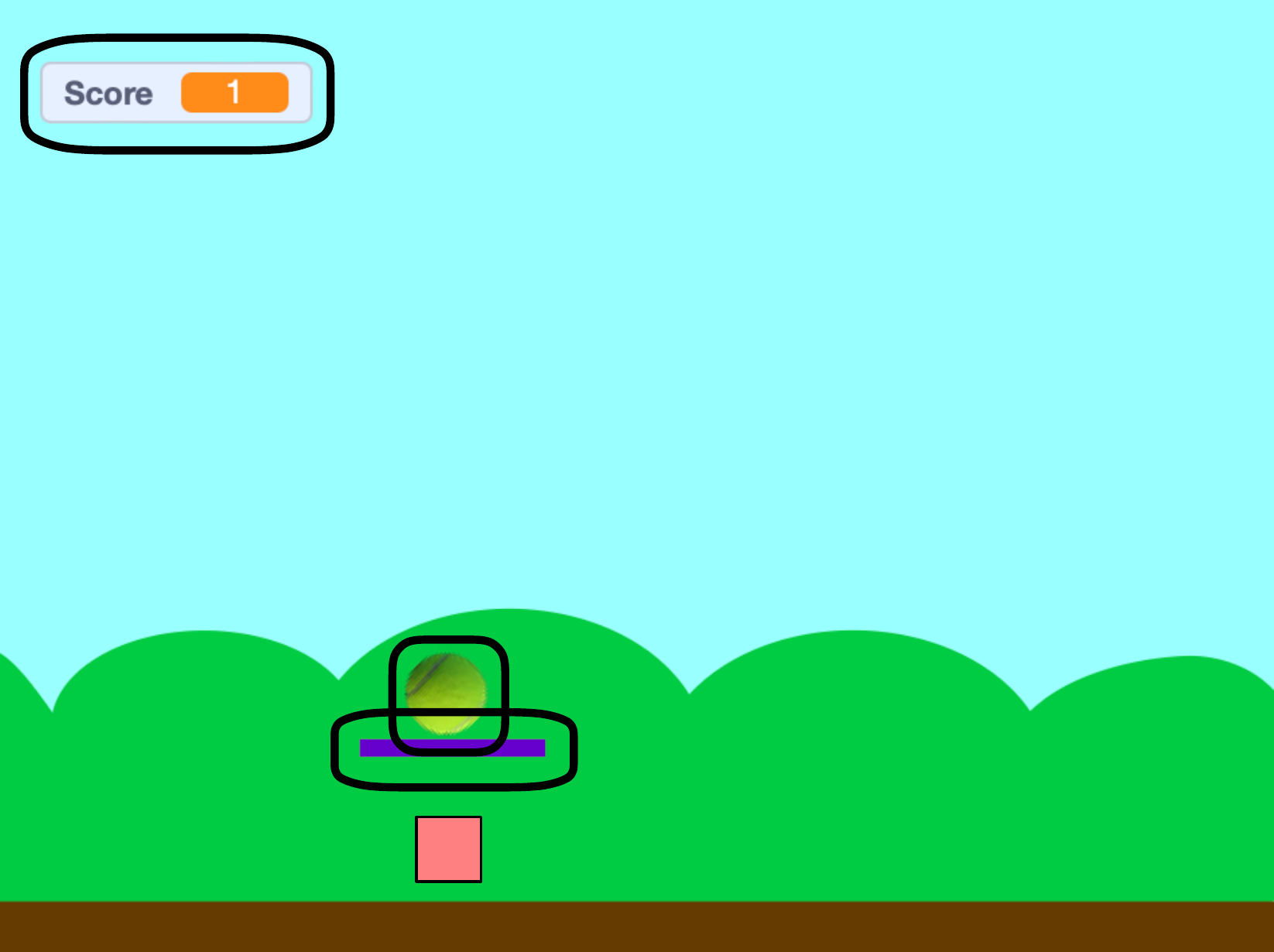}
    \caption{Player performs no action}
    \label{fig:Pong-3}
  \end{subfigure}%
    \hspace{.1em}
  \begin{subfigure}[b]{.49\columnwidth}
    \centering
    \includegraphics[width=\columnwidth]{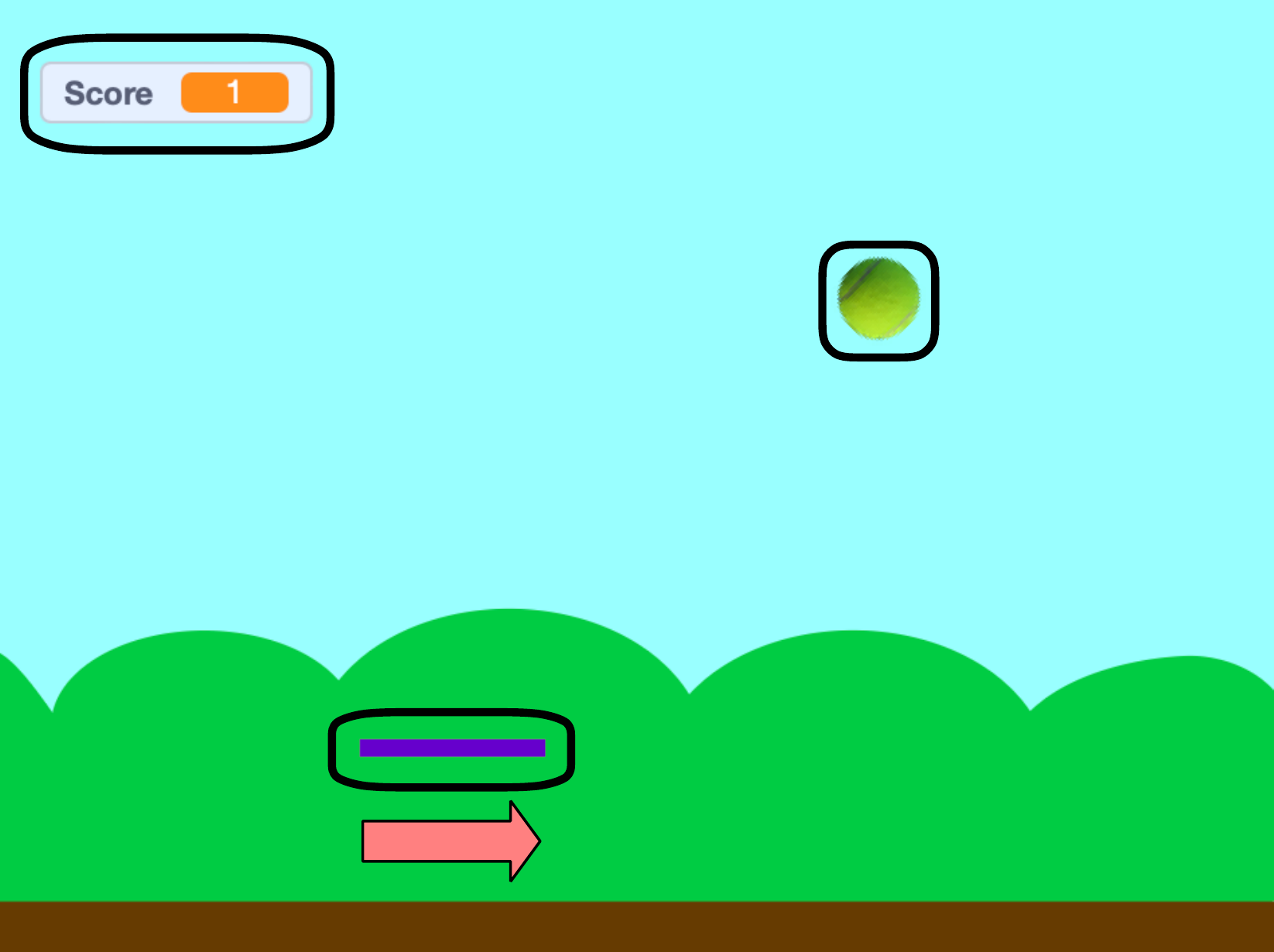}
    \caption{Player moves to the right}
    \label{fig:Pong-4}
  \end{subfigure}
  \caption{Four snapshots of a single gameplay trace. Training examples are derived from the figures' attributes (black squares) and mapped to labels of player actions (red arrow).}
  \label{fig:Pong}
  \vspace{-1.5em}
\end{figure}
% Context: Testing Games
%Since its inception, the video game industry has been growing continuously, reaching an approximate record revenue of 221 billion US dollars in 2023 and an expected annual growth rate of 6.25\% from 2023 to 2027~\cite{statista-videoGames}.
%In addition to the entertainment industry, games are increasingly incorporated into programming education because they spark and sustain student interest while introducing them to crucial programming concepts~\cite{fagerlund2021computational, topalli2018improving}.
%Furthermore, it has been shown that real-time formative feedback increases learning outcomes and self-efficacy~\cite{marwan2020adaptive, corbett2001locus}, encouraging instructors to develop unit tests for their students such that they can validate their programs~\cite{hovemeyer2013cloudcoder, edwards2017codeworkout}.
%However, despite the prevalence of games, they are usually still tested manually due to a lack of research in automated game testing~\cite{politowski2021survey}, forcing game developers, educators and students to test their games manually.
Generating adequate test input sequences for games is a daunting task, as games are designed to entertain players by challenging them with increasingly complex problems.
Conventional test generators based on fuzzing or evolutionary search~\cite{fraser2011evosuite, mao2016sapienz} are not able to play games in a meaningful way and therefore fail to reach advanced program states.
Even if a test generator manages to generate static input sequences that cover a game sufficiently, these input sequences are very likely flaky due to the inherent randomisation of games~\cite{gruber2021empirical}.
The \Neatest~\cite{feldmeier2022neuroevolution} approach tackles these challenges by generating dynamic tests in the form of neural networks that play games meaningfully.
These networks are optimised via neuroevolution to be robust against program randomisation and assembled in test suites where they serve as test input generators for games.

% Problem: Long training durations of evolutionary algorithms
%The evolutionary search applied in \Neatest may take a considerable amount of time, sometimes requiring long search durations to cover even relatively simple program statements.
\Cref{fig:Pong} shows a simple game, in which the player has to steer the paddle at the lower half of the screen to the left or right using the mouse as input device.
The objective of the game is to prevent the ball from touching the brown floor by moving the mouse such that the ball is juggled by the paddle.
Although this is a relatively simple task for a human player, neuroevolution requires a considerable amount of time to evolve networks capable of moving the mouse to the precise locations required for juggling the ball. 
%To make things worse, \Neatest needs to apply the evolution multiple times to create networks able to cover all source code statements and the corresponding behaviour.
The root cause of this inefficiency resides in the underlying algorithm (NEAT~\cite{stanley2002evolving}), which uses probabilistic mutation and crossover operators to co-evolve the weights and topologies of neural networks.
Due to the complex fitness landscape of evolving the architecture and weights of networks in conjunction and the probabilistic nature of the search operators, the evolution of networks that serve as test input generators even for a game as simple as the one shown in \cref{fig:Pong} may require several hours of network training.

% Idea: Learning by Viewing
In order to improve the evolution of test input-generating networks, we integrate information about human gameplay.
To achieve this, we extend the neuroevolution to mutate networks not only probabilistically but also systematically using conventional gradient descent based on traces of human gameplay (\emph{learning by viewing}). %\todo{Redundant introduction of the central idea: Here and in the next paragraph again.}
As demonstrated by \cref{fig:Pong}, these traces consist of snapshots that contain information about the game state, like the positions of figures or the states of variables (black squares), that led the player to perform an action, as well as the performed action itself (red arrow).
For instance, \cref{fig:Pong} shows four gameplay snapshots in which the player starts to move the mouse to the left toward the ball, then waits in the following two snapshots until the ball is caught, and finally moves to the right to catch the ball again.
%Such a dataset is established by observing humans while playing games and taking snapshots of the game states whenever they perform an action.
%For instance, considering the game shown in~\cref{fig:Pong}, the traces include information about the player actions, for example when the player moves the mouse to the left (\cref{fig:Pong-1} red arrow) in order to move the paddle in the direction of the falling ball, as well as information about the corresponding game state, like the position of figures or the state of variables (black squares) that led the player to perform an action. 
% Finally, we combine a set of many snapshots to establish a conventional supervised classification learning scheme, where the states of snapshots are used as network inputs and the respective player actions as class labels.
While it would take the evolution considerable time to achieve such gameplay, updating weights based on gameplay traces is comparatively quick.

%
% Summary Contribution
%We integrate gradient descent into the neuroevolution algorithm using an alternative mutation operator that updates the networks' weights based on gradient descent rather than random mutations. To provide the necessary training data, we propose a novel method of extracting training examples and classification labels from human gameplay such that they can be used for supervised learning.
In summary, we propose a novel method of extracting training data from human gameplay traces to improve the neuroevolutionary search for optimal network weights in the context of game testing with an alternative mutation operator based on gradient descent.
We implement this approach as an extension of the \Neatest framework~\cite{feldmeier2022neuroevolution} and conduct experiments on eight games written in \Scratch,  a block-based programming language designed for young learners.
% Contributions
In detail, the contributions of this paper are as follows:
\begin{itemize}
	\item We propose a way of recording human gameplay such that it can be used in a supervised learning scheme.
	\item We combine neuroevolution with gradient descent to improve the optimisation of networks that serve as test input generators for games.
	\item We implement the proposed approach as an extension to the publicly available game testing framework \Neatest.
	\item We empirically evaluate the approach on a set of 8 \Scratch games~\cite{maloney2010scratch} having varying complexities, sizes and genres.
\end{itemize}

Our experiments demonstrate that the integration of recorded gameplay traces successfully reduces the search time of evolving neural networks, enabling test input generators for games that achieve higher code coverage.

\section{Background}
This paper combines neuroevolution-based test input generation for games using search-based software testing methods with supervised network optimisation via gradient descent.

\subsection{Search-based Software Testing}
\label{section:sbst}
Search-based software testing (SBST)~\cite{mcminn2004search} uses meta-heuristic search algorithms to generate test inputs.
In this endeavour, the search, for instance in the form of an evolutionary algorithm, is governed by an objective function that measures how close a generated test is to reaching individual or all statements of a program.
Most of the time, the latter is realised using objective functions that combine the branch distance~\cite{korel1990automated} with the approach level~\cite{wegener2001evolutionary}.
The branch distance measures the distance between a generated test's last reached control dependency toward the target statement, whereas the approach level measures the number of control dependencies between the reached and targeted program statements.

Based on the chosen representation of viable solutions, search-based algorithms may be applied to a variety of software testing challenges, such as generating sequences of method calls~\cite{fraser2011evosuite, paolo2004evolutionary, baresi2010testful}, synthesising inputs for testing GUI applications~\cite{sell2019empirical, gross2012search, mao2016sapienz}, and optimising calls to REST services ~\cite{arcuri2019restful}.
All these synthesised test inputs are static in their behaviour as they cannot adapt to changes in program behaviour.
However, since randomised program behaviour is prevalent in video games, this lack of adaptability to behavioural changes limits the applicability of conventional search-based testing approaches to video games.
%Thus, Neuroevolution has been combined with SBST in a testing framework called \Neatest to generate test suites of neural networks capable of adapting to randomised program behaviour.

\subsection{Neuroevolution-based Test Generation}
\label{section:neatest}

\Neatest~\cite{feldmeier2022neuroevolution} is an extension of the \Whisker testing framework for \Scratch programs~\cite{stahlbauer2019testing} that tackles the many challenges of game testing by generating so-called dynamic test suites.
These test suites consist of neural networks, with each one of them serving as a test-input generator optimised to cover a targeted program statement regardless of the encountered and often randomised program behaviour.
In order to guarantee that the dynamically generated test inputs are \emph{robust} against heavy program randomisation, the framework does not add networks that manage to cover a statement once straight away to the final test suite.
Instead, these networks first have to pass a so-called \emph{robustness check} that tests whether a given network is able to reach the targeted program statement repeatedly in several randomised program executions.
 Only networks that pass this check are added to the final test suite as \emph{reliable} test cases.
 Whenever a statement is covered reliably, the search proceeds with the optimisation of networks for the next statement.

For a chosen target statement, \Neatest optimises networks via the neuroevolution algorithm NEAT~\cite{chen2006neuroevolution} that uses mutation and crossover operators over many generations to explore the search space of viable solutions.
Mutations introduce probabilistic variation by extending a network’s topological structure or by changing attributes of existing genes. 
Crossover forms a single child from two parents by combining the genes of both parents.
Target statements are selected by querying the control dependence graph (CDG)~\cite{CDG} for statements that are direct children of already covered program statements, which allows \Neatest to iteratively explore the game by seeding initial generations with prior solutions.
%\todo{This is a bit unclear since the approach level would always be with respect to a given test. Need to clarify this is with respect to all tests generated so far} 
%In case the received list of statements contains more than one element, \Neatest chooses the next statement randomly and continues with the optimisation process until the target is covered.
%To ensure that the search does not get stuck on a hard-to-reach program statement, a new statement may be selected if no progress has been made for several generations.
%At this point \Neatest selects the next target statement and optimises networks towards covering it reliably based on the \emph{NeuroEvolution of Augmenting Topologies} (NEAT) algorithm~\cite{stanley2002evolving}, which uses the in~\cref{section:sbst}
The search is guided by a combination of branch distance and approach level as objective function, such that \Neatest implicitly learns to win or lose games in different ways that reliably reach target statements.

By co-evolving the weights of networks together with their topology, NEAT relieves developers from the tedious process of finding suitable network architectures while still achieving good results in many application domains~\cite{risi2015neuroevolution, schmidhuber2007training, gomez2008accelerated}.
Due to flexible objective functions inherent to evolutionary algorithms, neuroevolution can be applied to many problem domains and may even be combined with other learning techniques~\cite{floreano2008neuroevolution}, such as gradient descent.

\subsection{Optimising Networks via Gradient Descent}
\label{section:gradient_descent}

Gradient descent combined with backpropagation~\cite{rumelhart1986learning} is the most common method of network training and optimises the weights of networks based on a dataset of training examples $X$.
Each of these examples is defined by a tuple $(x, y)$ consisting of a data point $x$ and a corresponding label $y$.
Given such a dataset and a neural network, the set of network weights $\Theta$ may be optimised by iterating over each training tuple $(x, y) \in X$ of the dataset and executing the following four steps on each data point:
First, in the so-called \emph{forward pass}, a given network is supplied with the training input $x$ to produce a prediction $\hat{y}$.
Based on the network prediction, we can determine how far the network is off from the correct prediction $y$ by computing the loss $L(y,\hat{y})$ via the problem-dependent loss function $L: \mathbb{R}x\mathbb{R} \rightarrow \mathbb{R}$.
Then, we conduct the \emph{backward pass} by computing the gradient $\nabla_{\Theta_i}$ of the loss function $L$ for each network weight $\Theta_i$ using the famous \emph{backpropagation} algorithm.
Finally, we update the network weights $\Theta_i$ via gradient descent by moving each weight value slightly in the direction of the negative gradient:
%\begin{equation}
$	\Theta_i = \Theta_i - \alpha * \nabla_{\Theta_i}$.
%\end{equation}
The learning rate $\alpha$ defines the size of the update step and is arguably the most crucial parameter of the gradient descent algorithm~\cite{goodfellow2016deep}: too small values lead to slow training progress, whereas too high values interfere with the optimisation progress as they may cause giant leaps within the landscape of the loss function.

Iterating over each training sample one by one is called \emph{true stochastic gradient descent} (SGD) and exposes an accurate albeit noisy optimisation landscape. 
Smooth landscapes may be obtained by combining all (\emph{batch gradient descent}) or several training examples (\emph{mini-batch stochastic gradient descent}) in so-called training batches and deferring the final weight update step until the combined gradient over a given training batch has been computed.
%Traditionally, networks are trained by conducting several passes, so-called epochs, over the entire dataset.
Since the networks are updated on the gradient derived from the loss function, the corresponding loss values keep shrinking over several epochs of iterating through the whole dataset.
Thus, the more the training dataset resembles the task one tries to solve, the better neural networks may be optimised via gradient descent. 
%Unfortunately, such training datasets for testing games are infeasible and non-existent due to the explosion of randomised program states.\todo{Unclear why that makes them infeasible}
We approximate a training dataset by recording human gameplay and combine gradient descent with evolutionary optimisation to improve the optimisation of test input-generating neural networks.
\section{Adapting Neural Networks to Human Gameplay Traces}
\label{section:records}
%In this work, we explore whether networks can be adapted to human gameplay traces such that they can serve as test input generators for randomised video games.
Our approach extends the  \Neatest~\cite{feldmeier2022neuroevolution} neuroevolution-based test generator for games with the gradient descent algorithm using traces of human gameplay as a training dataset.

\subsection{Synthesising Training Datasets from Human Gameplay Recordings}
We establish a dataset of training examples $D$ for the gradient descent algorithm by observing the actions a player takes when facing different program states while playing a game.
Whenever a player performs an input action, we take a snapshot $S=(x_i, y_i)$ that is defined by the tuple over the current game state $x_i$ and the executed player action $y_i$.
The training examples $x_i\in X$ are extracted from the set of feasible game states $X$, which may be infinitely large due to program randomisation.
On the other hand, classification labels $y_i\in Y$ are determined through the action a player executes and are bounded by the set of actions $Y$ for which corresponding event handlers exist in the game's source code.
All snapshots $n$ created within a single playthrough are then assembled into separate recording sessions $R_i=\{S_1,\dots,S_n\}$.
A recording session starts whenever the game is launched and ends when the player interrupts the execution of the game or when the game reaches a game over state. %\todo{I find this concept of recording sessions a bit confusing: First, I would have expected that snapshots are extracted from recording sessions, rather than assembling recording sessions from snapshots. Second, I would have expected a session to simply be a recording of everything a player did in one session. What's the purpose of grouping this into sessions?}
 %After a recording session has stopped, we check which program statements $p_i \in P$ of the game's source code $P$ have been reached by the current recording session.
Finally, we map each recording session $R_i$ to the set of covered program statements $P_{c_i} \subseteq P$ extracted from the game's source code $P$ and form a training dataset $D$ over all recording sessions: $D = \{(R_1, P_{c_1}), \dots, (R_m,P_{c_m})\}$.

\vspace{0.2cm}
\noindent\textbf{Data Points:} The training examples $x$ approximate the current game state and have to include all the information a neural network requires to make an informed decision.
Thus, which attributes are to be extracted from the game state depends on the respective programming environment. We define this approach for \Scratch games, but it easily generalises to other game genres and programming environments.
We iterate over each visible figure on the game screen and extract the following values that are bounded and normalised to $[-1, 1]$ based on value ranges imposed by \Scratch~\cite{maloney2010scratch}:
\begin{itemize}
\item \textbf{Position} defined by coordinates $a \in [-240, 240]$ and \\ $b\in[-180, 180]$ on a 2-dimensional game canvas.
\item \textbf{Heading direction} defined by an angle $\alpha \in [-180, 180]$ if a figure's rotation style is set to \emph{all around}.
\item \textbf{Costume} of a figure defined by the index $i \in N$ over the list of available costumes if the figure changes its appearance.
\item \textbf{Size} of a figure defined by the size (\%) of the selected costume. 
\item \textbf{Private variable} values $v \in \mathbb{R}$ if the hosted value is interpretable as a number.
\item \textbf{Distance} $d\in[-600, 600]$ to a sprite or colour if the figure contains listeners for touching other sprites or colours.
\end{itemize}
Besides attributes specific to visible figures, we also collect numeric values of variables ($v \in \mathbb{R}$) shared across all figures and the mouse position ($a \in [-240, 240]$, $b\in[-180, 180]$) if the game contains code that requires the player to use the mouse.

\vspace{0.2cm}
\noindent\textbf{Classification Labels:} Similar to the game states $X$, the set of available actions $Y$ is governed by the chosen programming environment, and even though we use \Scratch as an example environment, the presented approach is generalisable to other programming domains.
We group actions into two disjoint classes, \emph{keyboard actions} and \emph{mouse actions}.
%, based on the hardware used for executing the respective action.
Keyboard actions handle key presses on the keyboard by registering the timestamp $t$ and current game state $x_t$ whenever a key $k_i$ is being pressed.
As soon as the pressed key $k_i$ is released, we compute the duration of the key press $t_d$ by subtracting the saved timestamp $t$ from the current timestamp $t_c$ and create a snapshot $S_i = (x_t, y_t)$ where $y_t$ represents the key press action together with the press duration $y_t = (k_i, t_d)$.
Mouse actions consist of mouse clicks and mouse movements, with the former being recorded similarly to key presses.
To avoid an explosion of snapshots related to mouse movements, snapshots of mouse move actions are not created right away when the mouse begins to move.
Instead, we save the game state $x_t$ when the mouse starts moving, check periodically whether it still moves, and only create a snapshot $S_i$ if the movement stops for a predefined amount of steps.
The created snapshot $S_i = (x_t, y_t)$ consists of the saved game state $x_t$ and the final position of the mouse after it stopped moving $y_t = (a, b)$ with $a \in [-240, 240]$ and $b\in[-180, 180]$ restricted by the boundaries of the game canvas.
For instance, considering the gameplay trace in \cref{fig:Pong}, the first snapshot $S_i = (x_1, y_1)$ in which the player moves the mouse to the left (\cref{fig:Pong-1}) is defined by the snapshot's state $x_1$ when the player started moving the mouse, and the final position of the mouse $y_1$ when the paddle stopped (\cref{fig:Pong-2}).

\vspace{0.2cm}
\noindent\textbf{No-Operation Actions:}
%So far, we have established a dataset by recording all the actions and corresponding states a player can take while playing a game.
%However,
Sometimes it is also desirable in games to just wait for some events to happen without executing any actions.
For instance, in \cref{fig:Pong-2} and \cref{fig:Pong-3}, \mbox{the player does not perform any} actions for a couple of seconds and instead waits for the ball to drop down onto the paddle.
In order to capture such player behaviours, we include a no-operation action in the set of available actions $Y$, regardless of input event listeners defined by the game's source code.
These no-operation actions are identified by saving the game state $x_t$ and the current timestamp $t_0$ \mbox{whenever an action has been} executed.
Then, whenever another action is performed, we compare the time difference $t_{\mathit{diff}} = t_{\mathit{new}} - t_{0}$ \mbox{between the old $t_0$ and new action} $t_{\mathit{new}}$.
If $t_{\mathit{diff}}$ exceeds a predefined threshold value $\delta_t$, we record a no-operation action by generating a snapshot $S_i = (x_t, y_t)$ that is defined by the saved state $x_t$ at time step $t_0$ and the passed waiting time $y_t = t_{\mathit{diff}}$.
Since the maximum waiting duration $w_{\mathit{max}}$ during the evaluation of networks is bounded to avoid infinitely long no-operation actions, we periodically check if the maximum waiting duration was reached $t_{\mathit{diff}} \geq w_{\mathit{max}}$ and record the no-operation action together with the value of $w_{\mathit{max}}$ if this is the case.

\subsection{Combining Evolutionary Search with Gradient Descent-based Optimisation}

By combining neuroevolution with gradient descent-based weight optimisation, we seek to improve the exploitation phase of an evolutionary search.
Neuroevolution, with its probabilistic mutation operators, is responsible for exploring the fitness landscape and discovering points within the landscape that are near optima.
Based on these points, we use gradient descent as an improved exploitation step to increase the speed at which these points are optimised towards the near optima.
This speedup is twofold; by following the gradient we can optimise the network weights far more systematically  than by using probabilistic evolutionary operations.
Furthermore, gradient descent does not involve costly network evaluations due to the surrogate of recorded gameplay traces.

Adapting networks to human gameplay via gradient descent can only serve as a proxy for our true goal of optimising networks toward reliably covering targeted program statements.
Therefore, we are faced with two different fitness landscapes: the true landscape of generating reliable test input generators in the form of neural networks and the surrogate landscape that adapts networks to human gameplay using gradient descent.
However, we expect these two landscapes to have some overlap in the sense that optima in the surrogate correspond to points in the neighbourhood of optima within the true objective function.
%Due to the use of a surrogate, gradient descent may not optimise networks directly towards reaching selected target statements, but recent insights suggest that reaching global optima is usually not required to obtain good network performance~\cite{dauphin2014identifying, saxe2014exact, choromanska2015loss}.

We integrate gradient descent in \Neatest by combining NEAT's conventional weight mutation operator that changes the weights of a network probabilistically with conventional gradient descent-based weight optimisation.
Whenever the search algorithm has chosen to change a network's weights, we apply either gradient descent or the traditional weight mutation based on a predefined probability.
Thus, we refrain from discarding the mutation operation because our training dataset based on which gradient descent is performed serves only as a surrogate for our goal of testing games.
Furthermore, we may encounter program states that are either not covered by the recorded training dataset or only represented poorly.
For instance, statements related to losing a game may be covered accidentally but not systematically during a playthrough, ultimately causing overfitting when applying gradient descent.

If gradient descent has been chosen to optimise a network toward covering a selected target statement $p_t \in P$, we start by extracting a subset of recording sessions $R_{p_t} \subset R$ from our training dataset $D$, where each extracted session has covered the target statement $p_t$ during the corresponding playthrough.
Then based on the extracted snapshots $S_p$ of $R_p$, we apply true stochastic gradient descent (SGD) by executing the four steps outlined in \cref{section:gradient_descent} using the training examples $x_i$ and corresponding labels $y_i$ of a given snapshot $S_i \in S_{p_t}$ as the network input vector and desired network output.
SGD was chosen instead of mini-batch or full batch gradient descent since we sometimes only have a few data points for a given statement $p_t$, which mitigates the benefit of using batch sizes greater than one.

The gradient descent algorithm is further refined by two regularisation techniques.
First, we apply \emph{multitask learning}~\cite{caruana1993multitask} since our networks have to deal with the classification task of selecting an action \emph{and} the regression task of determining parameters for specific actions.
For instance, whenever a network decides to move the mouse, it also has to determine the coordinates to which the mouse should be moved.
Similar to our scenario, in multitask learning, a network is trained to perform multiple tasks at once given a single input $x_i$.
Networks trained in such learning environments benefit from improved generalisation capabilities due to parameter sharing~\cite{pham2018efficient}.
As loss functions, we use the \emph{categorical cross-entropy loss}~\cite{liu2016large} for selecting an action and the \emph{squared error loss}~\cite{ott2022joint} for predicting corresponding parameter values.

Besides multitask learning, we also apply \emph{early stopping}~\cite{bishop1995regularization, sjoberg1995overtraining} as a second regularisation technique in order to reduce the execution time of the gradient descent algorithm and increase the networks' generalisability by avoiding overfitting.
Early stopping is realised by extracting a validation split from our training dataset $S_p$  that is not used for computing gradients and updating the weights of our networks.
Instead, this validation split is only used to compute a corresponding validation error after each epoch of network training to observe the training progress.
If the validation error does not improve for a specifiable amount of consecutive epochs $\delta_e$, training stops, and the networks weights are set to the state when the validation error was the lowest.
However, as we have to deal with a sparse dataset, we can only create a validation split if we have sufficient training samples.
If this is not the case, we use the training loss over the entire dataset after each epoch to decide whether we should continue with training.
Thereby, early stopping no longer prevents overfitting but at least ensures that the training process is stopped if no more progress is made.

\section{Evaluation}
To evaluate whether the neuroevolution-based generation of test inputs for games can be improved by adapting networks to human gameplay, we aim to answer the following two research questions:
\begin{itemize}
    \item \textbf{RQ1 (Recordings):} What are properties of suitable gameplay recordings?
    \item \textbf{RQ2 (NE+SGD):} Does neuroevolution benefit from gradient descent-based optimisation?
\end{itemize}
Our proposed approach is implemented as an extension to the \mbox{\Neatest} algorithm, which is part of \Whisker~\cite{deiner2022automated}, an open-source test generator for \Scratch programs.
The experiment dataset, parameter configurations, raw results of the experiments, and scripts for reproducing the results are publicly available on Figshare: 
\begin{center}
	\url{https://figshare.com/articles/dataset/22068875}
\end{center}

\subsection{Dataset}
\label{section:dataset}

\begin{figure*}
\begin{minipage}[t]{0.12\textwidth}
\includegraphics[width=\textwidth]{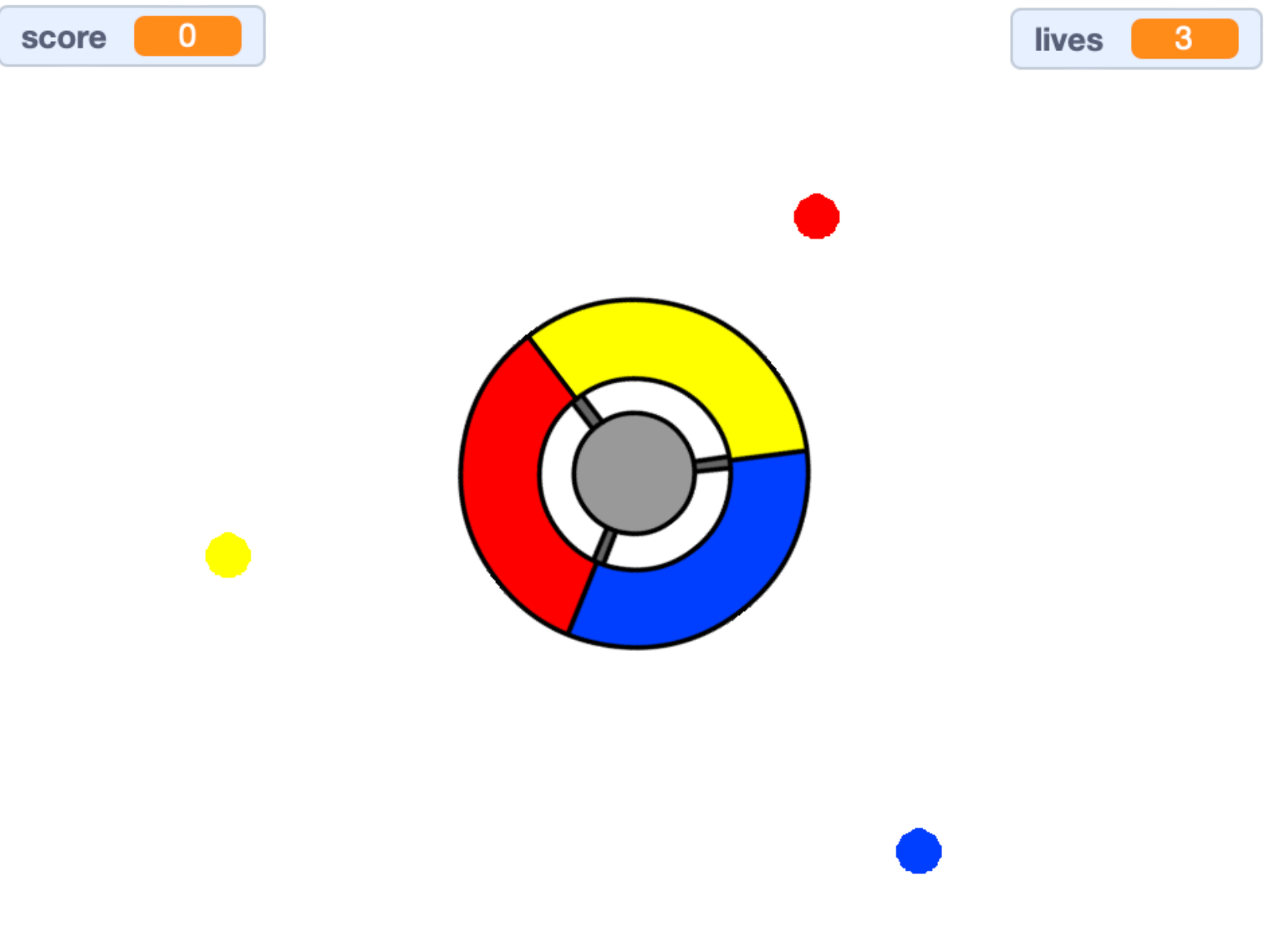}
\caption*{CatchTheDots}
\label{fig:Dataset}
\end{minipage}
\begin{minipage}[t]{0.12\textwidth}
\includegraphics[width=\textwidth]{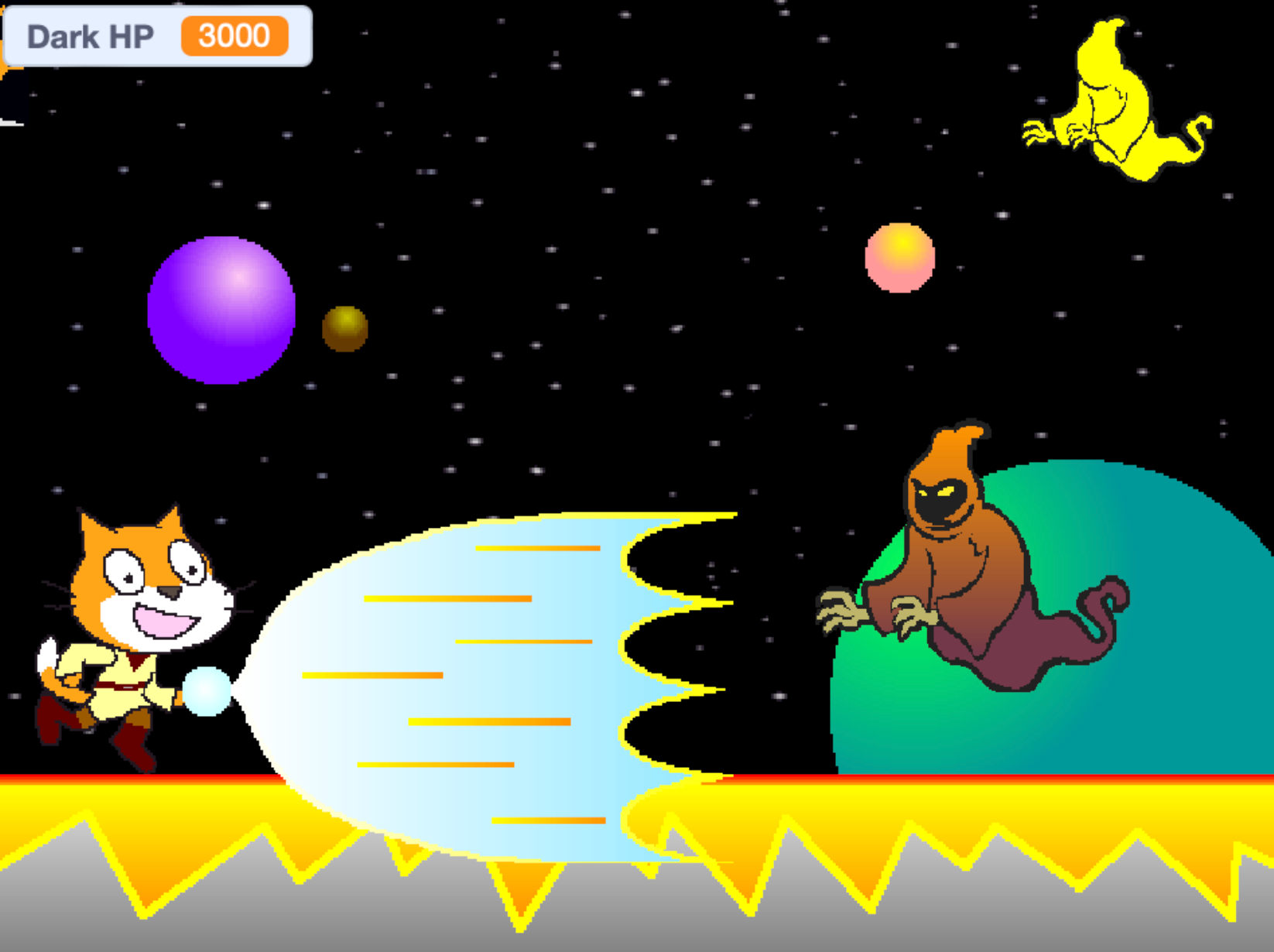}
\caption*{FinalFight}
\end{minipage}
\begin{minipage}[t]{0.12\textwidth}
\includegraphics[width=\textwidth]{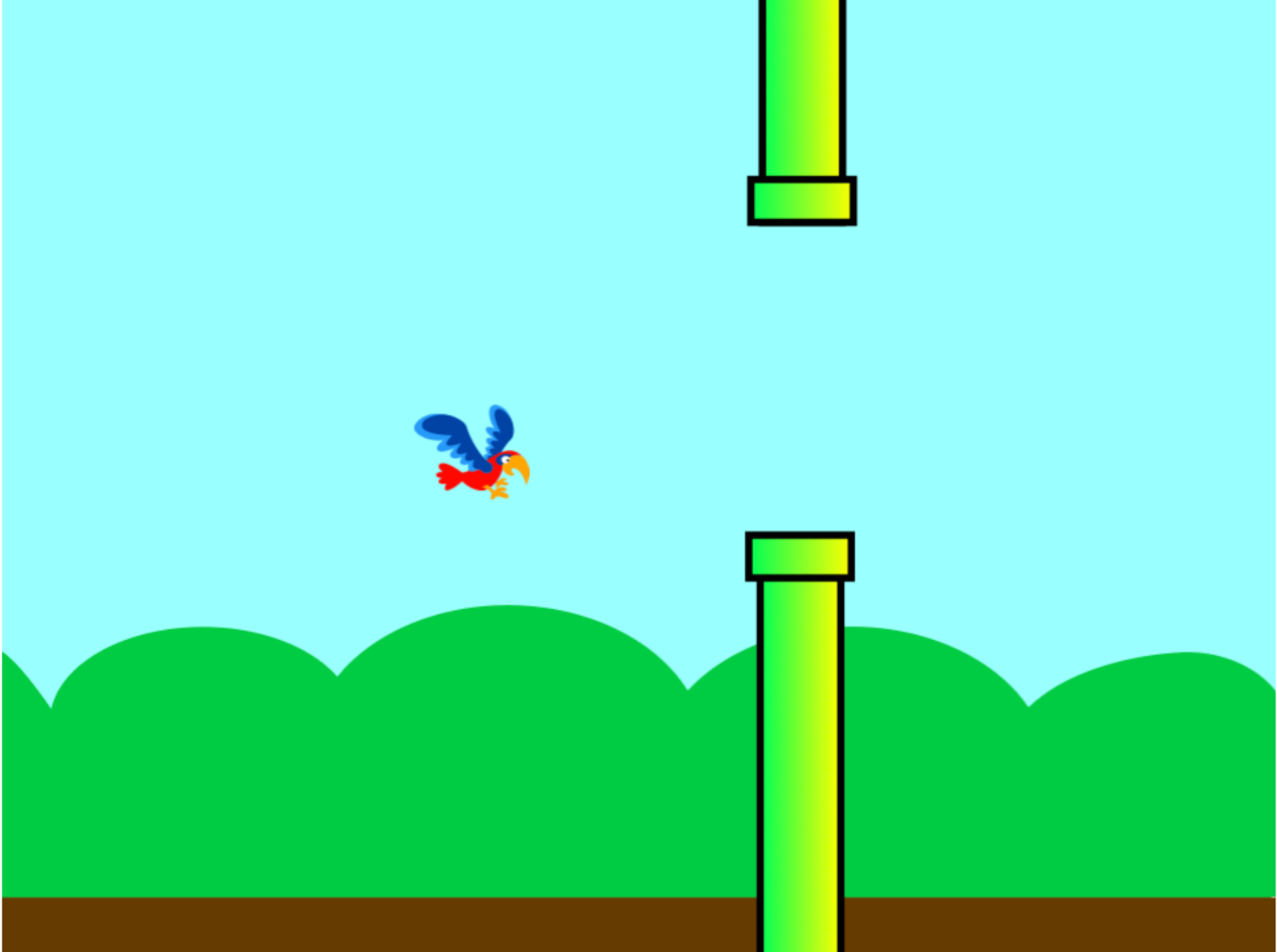}
\caption*{FlappyParrot}
\end{minipage}
\begin{minipage}[t]{0.12\textwidth}
\includegraphics[width=\textwidth]{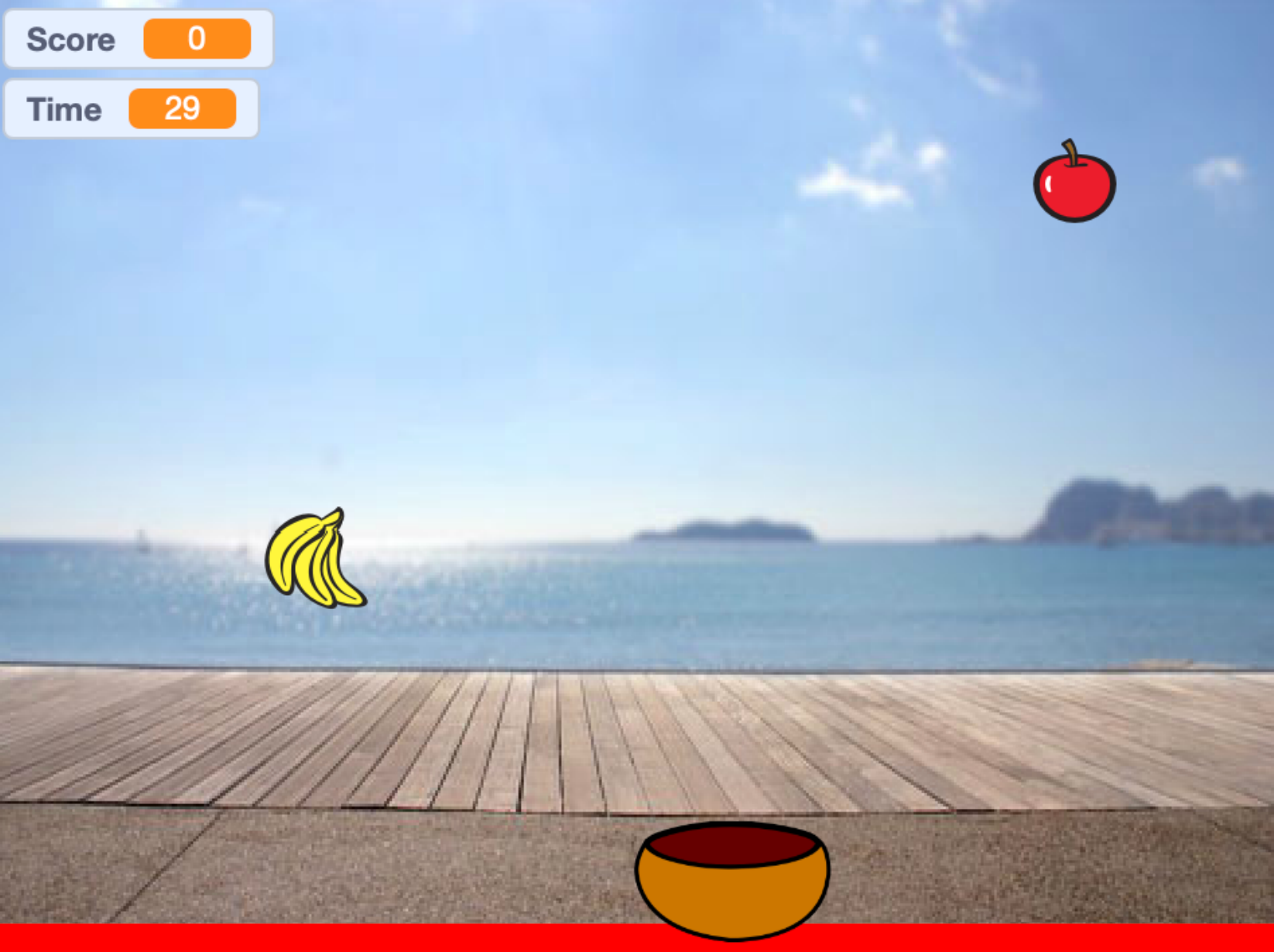}
\caption*{FruitCatching}
\end{minipage}
\begin{minipage}[t]{0.12\textwidth}
\includegraphics[width=\textwidth]{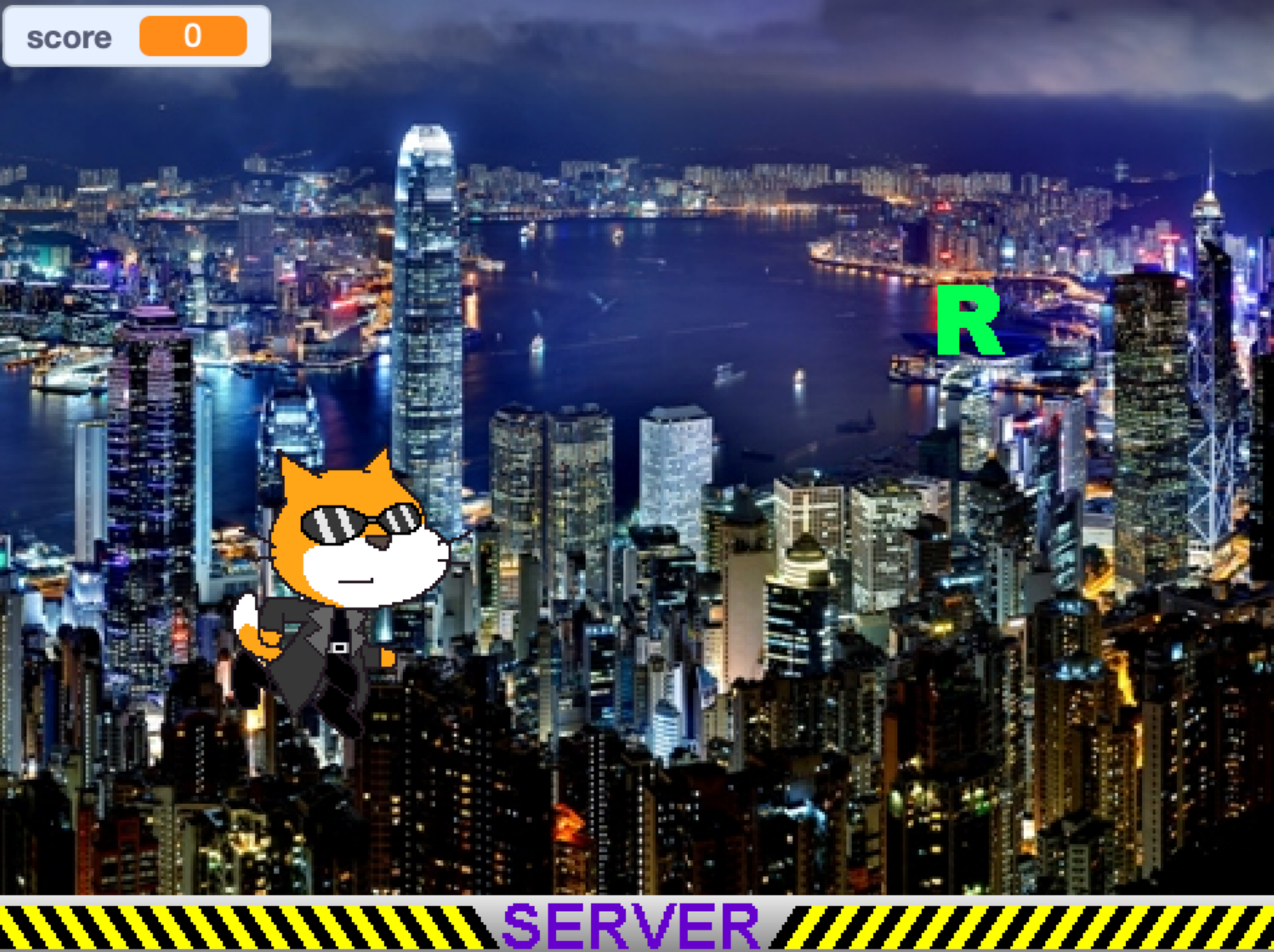}
\caption*{HackAttack}
\end{minipage}
\begin{minipage}[t]{0.12\textwidth}
\includegraphics[width=\textwidth]{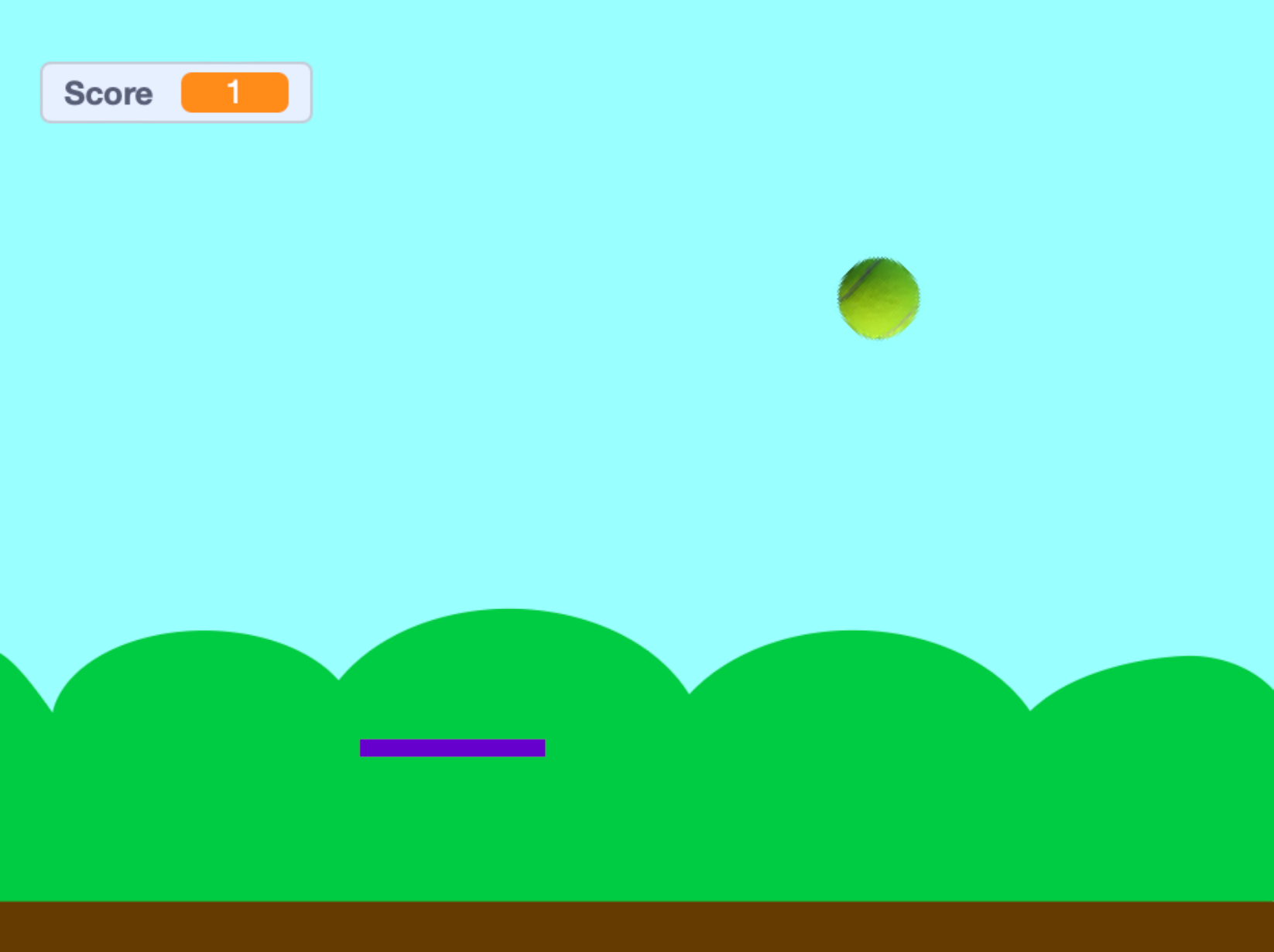}
\caption*{Pong}
\end{minipage}
\begin{minipage}[t]{0.12\textwidth}
\includegraphics[width=\textwidth]{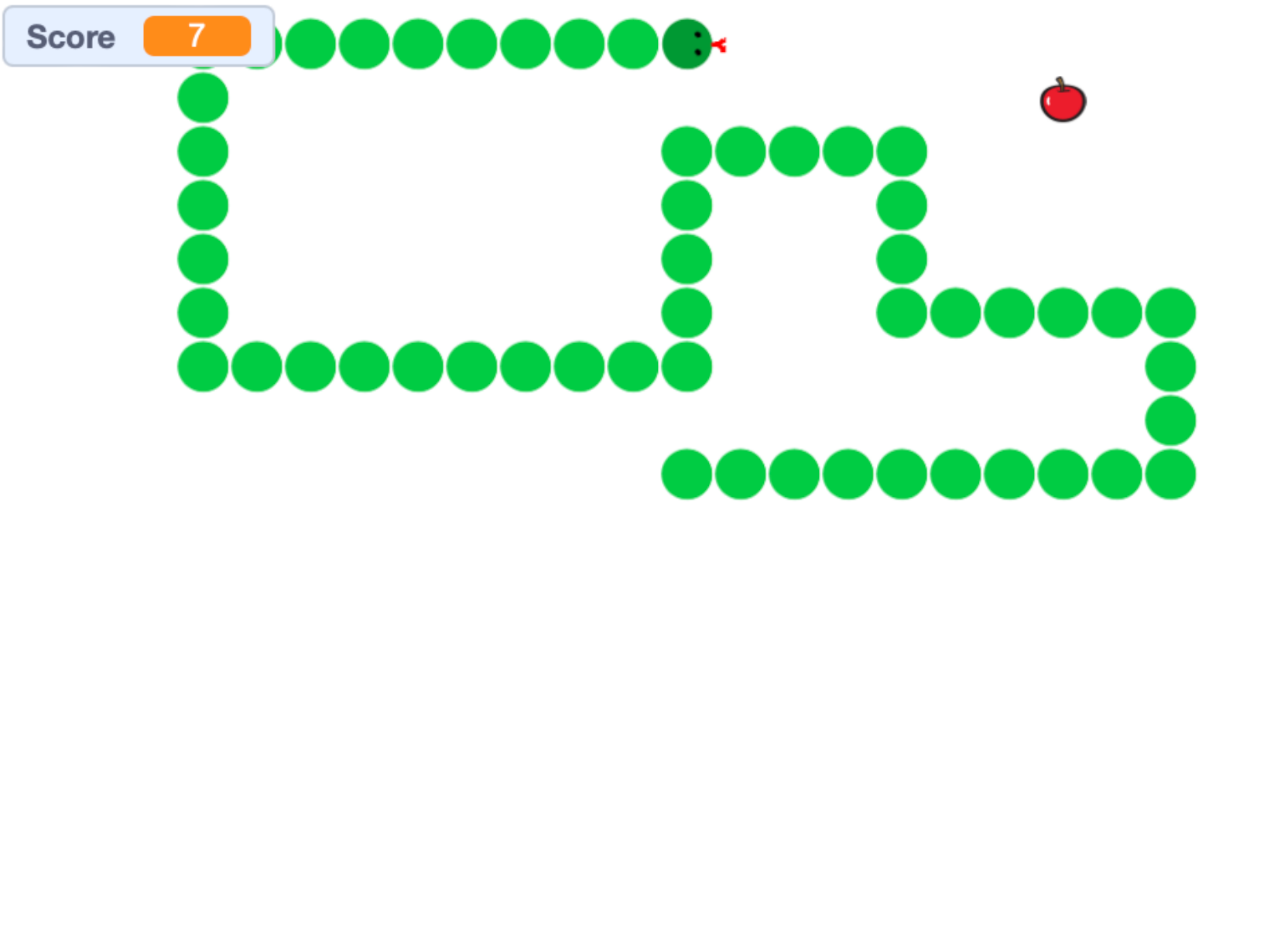}
\caption*{Snake}
\end{minipage}
\begin{minipage}[t]{0.12\textwidth}
\includegraphics[width=\textwidth]{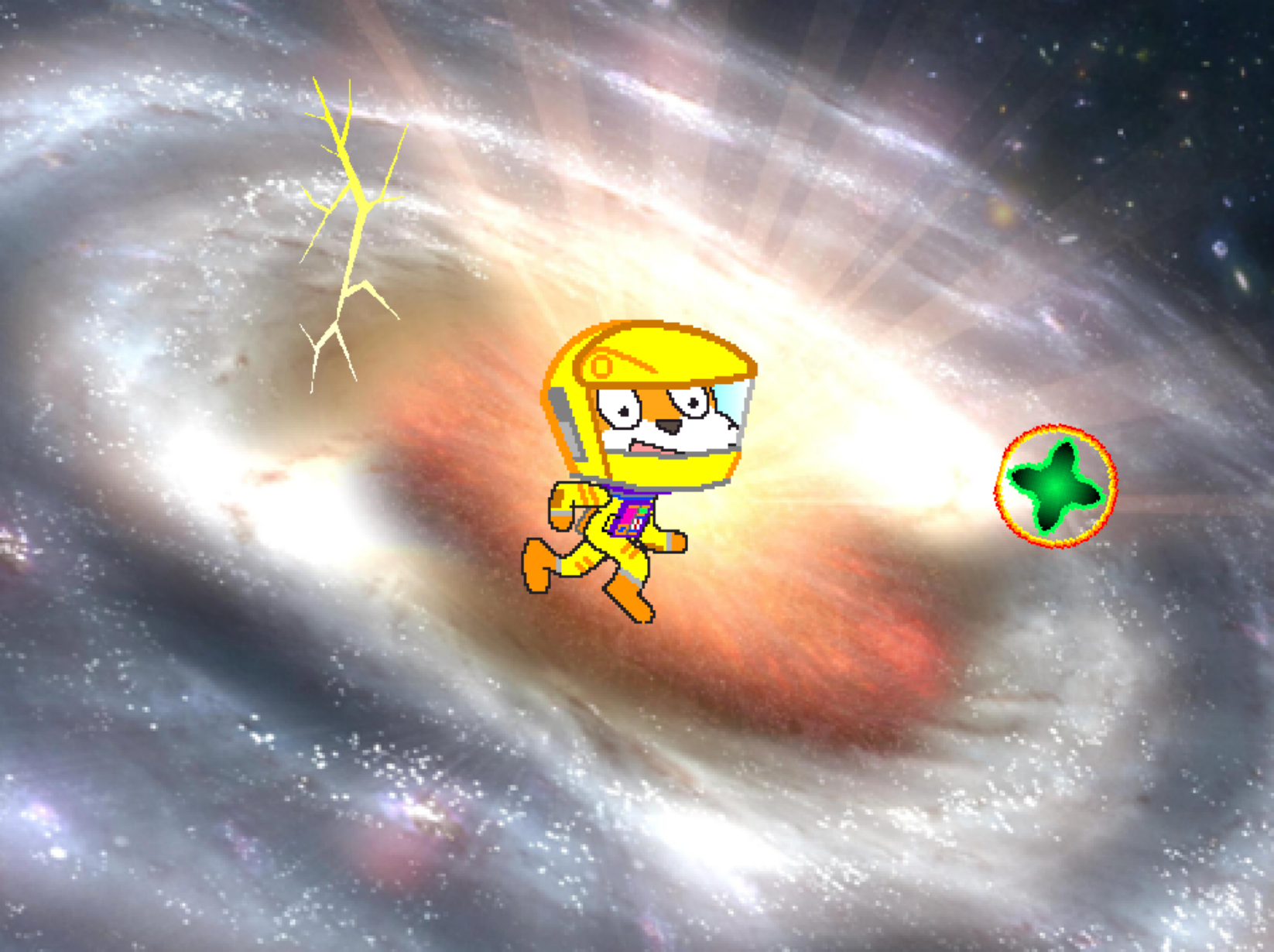}
\caption*{SpaceOdyssey}
\end{minipage}
\caption{The eight dataset games.}
\label{fig:Dataset}
\end{figure*}

\begin{table}[tb]
    \centering
    \caption{Evaluation dataset}
    \label{tab:Dataset}
        \vspace{-1em}
    \resizebox{\columnwidth}{!}{
        \begin{tabular}{lrrrlrrr}
            \toprule
            Project & \rotatebox{90}{\texttt{\#} Figures} & \rotatebox{90}{\texttt{\#} Scripts} & \rotatebox{90}{\texttt{\#} Statements} &
            Project & \rotatebox{90}{\texttt{\#} Figures} & \rotatebox{90}{\texttt{\#} Scripts} & \rotatebox{90}{\texttt{\#} Statements} \\
            \midrule
            CatchTheDots       & 4  & 10 & 82  & HackAttack    & 6   & 19   & 93    \\
            FinalFight         & 13 & 48 & 286 & Pong          & 2   & 2    & 15   \\
            FlappyParrot       & 2   & 7 & 37  & Snake         & 3   & 14   & 60    \\
            FruitCatching      & 3   & 4 & 55  & SpaceOdyssey & 4   & 13   & 116    \\
            \bottomrule
        \end{tabular}
    }
            \vspace{-1em}

\end{table}

We reuse the dataset established in the evaluation of \Neatest~\cite{feldmeier2022neuroevolution}, consisting of \Scratch programs extracted from introductory books, tutorials and prior work~\cite{stahlbauer2019testing}, as this resembles the target group of \Scratch; young children who learn to program.
The dataset contains 25 games filtered from a total of 187 programs by removing projects that are not games, trivial games and games without randomisation.
For our evaluation, we apply three additional filters to increase the expressiveness of our results.
First, we exclude easy games where \Neatest covers all program statements within one hour, as further speedups are not required for these games.
Second, since we are interested in analysing whether human gameplay traces can improve the search toward reaching advanced program states, we focus on games that can be won within a reasonable amount of time. %\todo{Why? Is this because we want to check a winning state, or because we want to be able to include training data that represents winning, and can only do that if we ourselves can succeed?}
Finally, aligning with the goal of our approach to reduce the search time by improving exploitation via gradient descent, we remove games with statements that are entirely dependent on randomness and not on compelling gameplay.
Applying these three additional filters results in the eight \Scratch games depicted in~\cref{fig:Dataset}, which involve many different game genres and ways of interacting with the program.
As can be seen in \cref{tab:Dataset}, the evaluation subjects have an average program size of 93 statements and involve up to 13 different figures. Note that \Scratch is a domain-specific language, such that only a few statements (blocks) are needed to implement complex and challenging game behaviour, resulting in smaller programs.

\subsection{Methodology}
All experiments were conducted on a dedicated computing cluster of nine AMD EPYC 7443P CPU nodes running on 2.85 GHz.
To reduce the required time for network evaluations, we set the integrated acceleration factor of \Whisker to a factor of 10, which accelerates program executions without introducing deviating program behaviour~\cite{deiner2022automated}.
Following the approach of \Neatest explained in \cref{section:neatest}, a target is only considered covered if a network manages to cover the same statement repeatedly in 10 randomised program executions.
%Note, that since we are dealing with games implemented in the block-based programming language \Scratch, program statements are essentially represented by building blocks that can be combined in meaningful ways to implement the desired program logic.
Together with a population size of 100 networks and a search duration of five hours, we ensure a reasonable amount of exploration and exploitation while retaining a convenient search duration.
Early stopping terminates the gradient descent algorithm after 30 epochs without progress ($\delta_e = 30$), as further improvements are very unlikely past this threshold.
Finally, we deactivated the average weight difference in NEAT’s compatibility computation since this would potentially lead to an explosion of species due to large weight changes induced by gradient descent.
%due to powerful weight changes induced by gradient descent, we discard the average weight difference in NEAT's compatibility computation.
The remaining NEAT parameters are set according to prior work~\cite{stanley2002evolving}.
\begin{table}[tb]
    \centering
    \caption{Number of training examples (and no-operation proportions) of the in RQ1 analysed gameplay traces.}
    \label{tab:Records}
        \vspace{-1em}
    \resizebox{\columnwidth}{!}{
        \begin{tabular}{lrrrrr}
            \toprule
            Program         & $\delta_t$=10                                                             & $\delta_t$=100                                                                    & $\delta_t$=$\infty$/3m                                                   & 1m                                                                   & 30s                                                                      \\
            \midrule
            CatchTheDots    & \SamplesWaitTenCatchTheDots (\WaitPropWaitTenCatchTheDots)    & \SamplesWaitHundredCatchTheDots (\WaitPropWaitHundredCatchTheDots)    & \SamplesWaitNoneThreeMinCatchTheDots (\WaitPropWaitNoneThreeMinCatchTheDots)      & \SamplesOneMinCatchTheDots (\WaitPropOneMinCatchTheDots)      & \SamplesThirtySecCatchTheDots (\WaitPropThirtySecCatchTheDots)    \\
            FinalFight      & \SamplesWaitTenFinalFight (\WaitPropWaitTenFinalFight)        & \SamplesWaitHundredFinalFight (\WaitPropWaitHundredFinalFight)        & \SamplesWaitNoneThreeMinFinalFight (\WaitPropWaitNoneThreeMinFinalFight)          & \SamplesOneMinFinalFight (\WaitPropOneMinFinalFight)          & \SamplesThirtySecFinalFight (\WaitPropThirtySecFinalFight)        \\
            FlappyParrot    & \SamplesWaitTenFlappyParrot (\WaitPropWaitTenFlappyParrot)    & \SamplesWaitHundredFlappyParrot (\WaitPropWaitHundredFlappyParrot)    & \SamplesWaitNoneThreeMinFlappyParrot (\WaitPropWaitNoneThreeMinFlappyParrot)      & \SamplesOneMinFlappyParrot (\WaitPropOneMinFlappyParrot)      & \SamplesThirtySecFlappyParrot (\WaitPropThirtySecFlappyParrot)    \\
            FruitCatching   & \SamplesWaitTenFruitCatching (\WaitPropWaitTenFruitCatching)  & \SamplesWaitHundredFruitCatching (\WaitPropWaitHundredFruitCatching)  & \SamplesWaitNoneThreeMinFruitCatching (\WaitPropWaitNoneThreeMinFruitCatching)    & \SamplesOneMinFruitCatching (\WaitPropOneMinFruitCatching)    & \SamplesThirtySecFruitCatching (\WaitPropThirtySecFruitCatching)  \\
            HackAttack      & \SamplesWaitTenHackAttack (\WaitPropWaitTenHackAttack)        & \SamplesWaitHundredHackAttack (\WaitPropWaitHundredHackAttack)        & \SamplesWaitNoneThreeMinHackAttack (\WaitPropWaitNoneThreeMinHackAttack)          & \SamplesOneMinHackAttack (\WaitPropOneMinHackAttack)          & \SamplesThirtySecHackAttack (\WaitPropThirtySecHackAttack)        \\
            Pong            & \SamplesWaitTenPong (\WaitPropWaitTenPong)                    & \SamplesWaitHundredPong (\WaitPropWaitHundredPong)                    & \SamplesWaitNoneThreeMinPong (\WaitPropWaitNoneThreeMinPong)                      & \SamplesOneMinPong (\WaitPropOneMinPong)                      & \SamplesThirtySecPong (\WaitPropThirtySecPong)                    \\
            Snake           & \SamplesWaitTenSnake (\WaitPropWaitTenSnake)                  & \SamplesWaitHundredSnake (\WaitPropWaitHundredSnake)                  & \SamplesWaitNoneThreeMinSnake (\WaitPropWaitNoneThreeMinSnake)                    & \SamplesOneMinSnake (\WaitPropOneMinSnake)                    & \SamplesThirtySecSnake (\WaitPropThirtySecSnake)                  \\
            SpaceOdyssey    & \SamplesWaitTenSpaceOdyssey (\WaitPropWaitTenSpaceOdyssey)    & \SamplesWaitHundredSpaceOdyssey (\WaitPropWaitHundredSpaceOdyssey)    & \SamplesWaitNoneThreeMinSpaceOdyssey (\WaitPropWaitNoneThreeMinSpaceOdyssey)      & \SamplesOneMinSpaceOdyssey (\WaitPropOneMinSpaceOdyssey)      & \SamplesThirtySecSpaceOdyssey (\WaitPropThirtySecSpaceOdyssey)    \\
            \bottomrule
        \end{tabular}
    }
            \vspace{-1em}
\end{table}

\vspace{0.2cm}
\noindent\textbf{RQ1 (Recordings):} In our first research question, we analyse how different recording properties affect the optimisation goal of evolving reliable network test generators.
We start by evaluating the effects of different no-operation thresholds $\delta_t \in [10, 100, \infty]$, which govern, as shown in \cref{tab:Records}, how many of these operations are included in recorded gameplay traces.
A second experiment analyses three different recording durations (30 seconds, 1 minute and 3 minutes), which determine the number of data points in the training dataset.
Since the proportion of no-operation actions behaves approximately linear to the recording duration, we start by evaluating the threshold $\delta_t$ and continue in the second experiment with the best-performing threshold value.

The learning rate of the gradient descent algorithm (\cref{section:gradient_descent}) is set to a fixed value of 0.1 for all experiments since this value achieved the best results for all examined wait thresholds $\delta_t$ in preliminary experiments.
Furthermore, all game traces were recorded by a single player who knew the games beforehand and focused on playing as well as possible. 
To avoid side effects induced by the evolutionary weight operation, we deactivated the weight mutation operator for both experiments such that weights can only be modified via gradient descent.
RQ1 is answered by comparing the obtained results in both experiments based on the achieved coverage over time and the average coverage reached after the search budget of five hours has been exhausted. 
Furthermore, to evaluate whether the analysed recordings are able to guide the search toward reaching challenging program statements, we manually identified for each program statements that represent a winning state and report how often these winning states are reached within the 30 experiment repetitions.
Finally, to mitigate the effects of randomisation inherent to games and the network optimisation process, each configuration is run 30 times per project.

\vspace{0.1cm}
\noindent\textbf{RQ2 (NE+SGD):} The second research question examines whether the neuroevolution-based input generation benefits from adapting networks to human gameplay traces.
To this end, we select the best-performing recordings of RQ1 and compare different probabilities (0\%, 30\%, 60\%, 100\%) of applying gradient descent instead of NEAT's conventional weight mutation~\cite{stanley2002evolving}.
Besides the 0\% baseline, which represents the original \Neatest approach without gradient descent, as a second reference point we include a random test generator that creates sequences of static test inputs by randomly choosing actions from the same set of processable user inputs \Neatest uses to choose an action from~\cite{deiner2022automated}.
Similar to the first research question, we repeat the experiment for each configuration and game 30 times and compare the achieved coverages and the number of reached winning states.
Furthermore, we determine statistical significance using the \emph{Mann-Whitney-U-test}~\cite{Mann-Whitney-U} by comparing the gradient descent probabilities of 30\%, 60\% and 100\% against the 0\% baseline.

\subsection{Threats to Validity}
\textbf{External Validity:} The evaluation dataset consists of 8 \Scratch games with varying genres, complexity and ways players can interact with them.
However, we cannot guarantee that the results generalise well to other \Scratch programs or games implemented in different programming environments.

\vspace{0.1cm}

\noindent \textbf{Internal Validity:} Heavy program randomisation in games and the inherent probabilistic operations of evolutionary algorithms may cause varying experiment outcomes.
Thus, we mitigate the effects of randomisation by repeating each experiment 30 times and exclude \Scratch programs in which sophisticated gameplay is not the main requirement for reaching advanced program states.
In our evaluation, a single player recorded all game traces, but different play styles may lead to different results.

\vspace{0.1cm}

\noindent \textbf{Construct Validity:}
Experiment results are compared based on the achieved block coverage, which is the equivalent of statement coverage in text-based programming languages.
However, block coverage must be treated with caution because large parts of \Scratch programs can often already be covered by executing arbitrary actions.
Thus, we excluded games from the dataset that do not demand meaningful gameplay to reach advanced program states.

\subsection{RQ1 (Recordings): What are properties of suitable gameplay recordings?}
\label{section:rq1}
\begin{figure}[!tbp]
  \begin{subfigure}[b]{.5\columnwidth}
    \centering
    \includegraphics[width=\columnwidth]{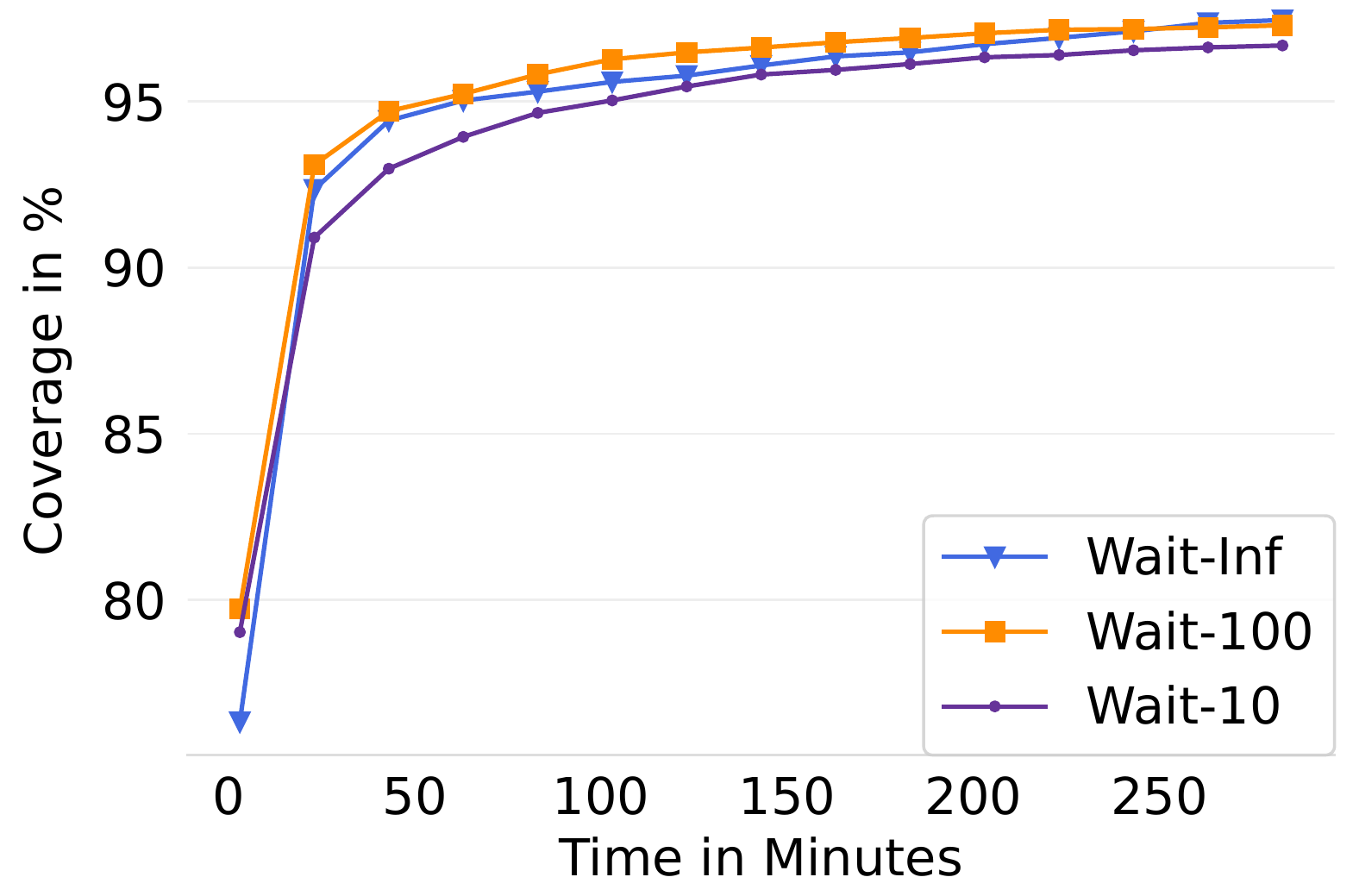}
    \caption{No-operation thresholds $~\delta_t$}
    \label{fig:RQ1-Waits}
  \end{subfigure}%
  \begin{subfigure}[b]{.5\columnwidth}
      \centering
    \includegraphics[width=\columnwidth]{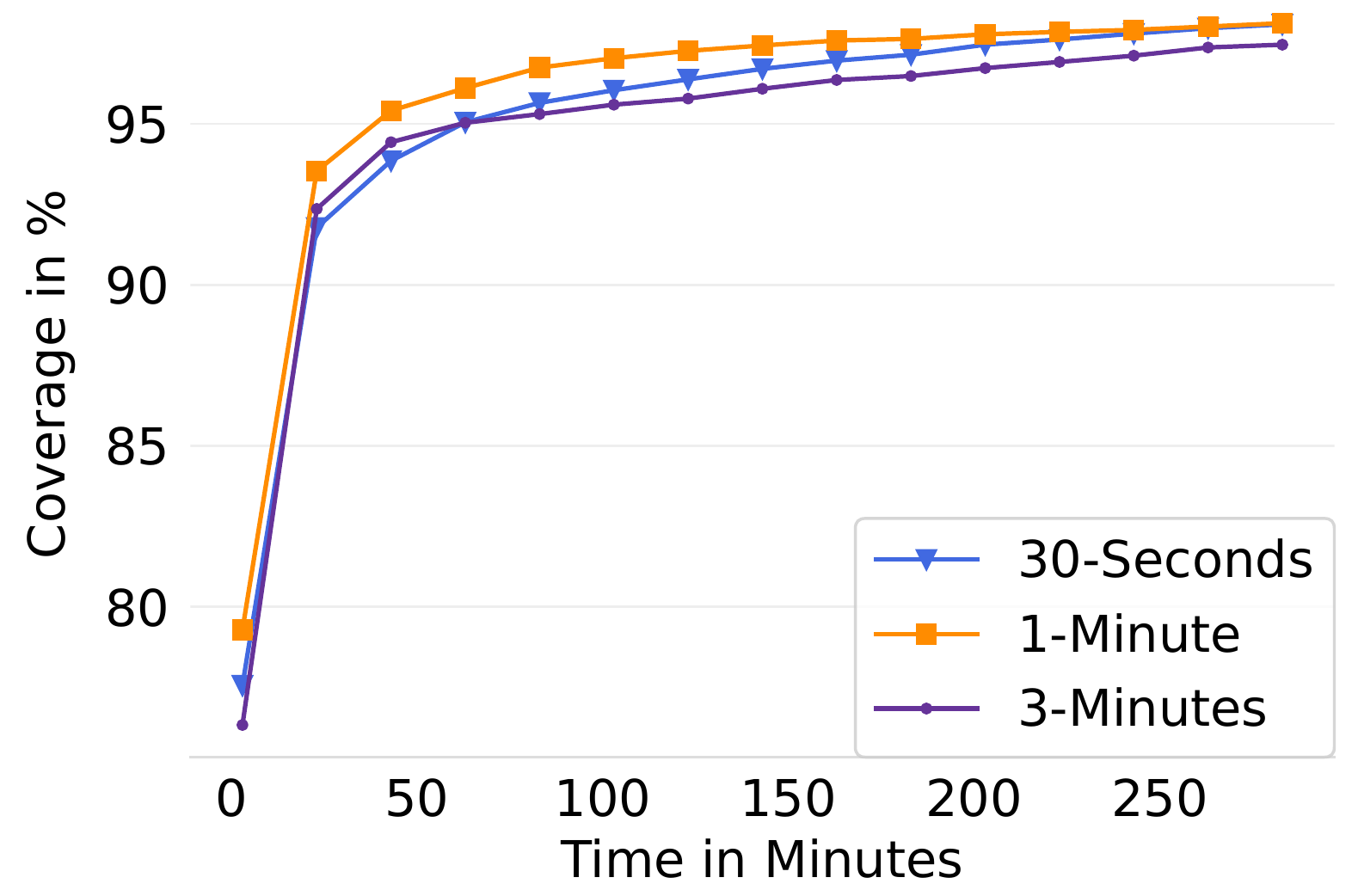}
    \caption{Recording durations}
    \label{fig:RQ1-Sizes}
  \end{subfigure}
  \caption{Average coverage over time across all games using varying no-operation thresholds and recording durations.}
\end{figure}

The first research question investigates how no-operation actions and the total number of training examples affect the evolution of test input generating neural networks.
\cref{tab:RQ1} shows that the inclusion of no-operation actions seems to impair the evolution process since an infinite threshold of $\delta_t$ that discards no-operation actions entirely (see \cref{tab:Records}) achieves the best overall coverage of \TotalMeanCovWaitInf\%, and the most winning states with an average of \WinningStatesAverageWaitInf/30 runs in which winning states were reached across all games.
\Cref{fig:RQ1-Waits} demonstrates this observation once more as the low threshold of $\delta_t=10$ achieves less coverage in the same amount of time than the higher thresholds.
These results are counterintuitive as they suggest an unbalanced training dataset in which the networks are discouraged from performing no-operation actions, even though they may be beneficial in some scenarios.
We believe no-operation actions cause harm to the search progress because they depend on more than just the state of a single snapshot.
In fact, their duration is governed by the game states that pass by between two explicit player actions.
One way to include this information would be to extend the proposed approach to process sequences of snapshots, for instance, by explicitly evolving recurrent neural networks~\cite{elsaid2019evolving}.

\begin{table}[tb]
    \centering
    \caption{Average coverage (C) and reached winning states (W) of recordings with different no-operation thresholds ($\delta_t$) and recording durations in minutes (m) and seconds (s).}
    \setlength\tabcolsep{.75\tabcolsep}
    \label{tab:RQ1}
        \vspace{-1em}
    \resizebox{\columnwidth}{!}{
    \setlength{\tabcolsep}{2pt} 
        \begin{tabular}{lrr@{\extracolsep{5pt}}rr@{\extracolsep{5pt}}rr@{\extracolsep{5pt}}rr@{\extracolsep{5pt}}rr}
            \toprule
            \multicolumn{1}{r}{}        & \multicolumn{2}{c}{$\delta_t$=10}                       & \multicolumn{2}{c}{$\delta_t$=100}                                           & \multicolumn{2}{c}{$\delta_t$=$\infty$/3m}                           & \multicolumn{2}{c}{1m}                                                  & \multicolumn{2}{c}{30s}  \\
                                                  \cline{2-3}                                       \cline{4-5}                                                                 \cline{6-7}                                                             \cline{8-9}                                                                 \cline{10-11}
            Program         &                       C  &  W                                       &   C &  W                                                                    &              C          &  W                                          &        C   &  W                                                         &  C          &  W                                                                \\
            \midrule
            CatchTheDots    & \MeanCovWaitTenCatchTheDots & \WinningStatesWaitTenCatchTheDots    & \MeanCovWaitHundredCatchTheDots& \WinningStatesWaitHundredCatchTheDots    & \MeanCovWaitInfCatchTheDots &\WinningStatesWaitInfCatchTheDots      & \MeanCovOneMinCatchTheDots &\WinningStatesSizeOneminCatchTheDots      & \MeanCovThirtySecCatchTheDots &\WinningStatesSizeThirtysecCatchTheDots    \\
            FinalFight      & \MeanCovWaitTenFinalFight&  \WinningStatesWaitTenFinalFight        & \MeanCovWaitHundredFinalFight &\WinningStatesWaitHundredFinalFight        & \MeanCovWaitInfFinalFight &\WinningStatesWaitInfFinalFight          & \MeanCovOneMinFinalFight &\WinningStatesSizeOneminFinalFight          & \MeanCovThirtySecFinalFight& \WinningStatesSizeThirtysecFinalFight        \\
            FlappyParrot    & \MeanCovWaitTenFlappyParrot&  \WinningStatesWaitTenFlappyParrot    & \MeanCovWaitHundredFlappyParrot &\WinningStatesWaitHundredFlappyParrot    & \MeanCovWaitInfFlappyParrot &\WinningStatesWaitInfFlappyParrot      & \MeanCovOneMinFlappyParrot &\WinningStatesSizeOneminFlappyParrot      & \MeanCovThirtySecFlappyParrot &\WinningStatesSizeThirtysecFlappyParrot    \\
            FruitCatching   & \MeanCovWaitTenFruitCatching & \WinningStatesWaitTenFruitCatching  & \MeanCovWaitHundredFruitCatching& \WinningStatesWaitHundredFruitCatching  & \MeanCovWaitInfFruitCatching& \WinningStatesWaitInfFruitCatching    & \MeanCovOneMinFruitCatching& \WinningStatesSizeOneminFruitCatching    & \MeanCovThirtySecFruitCatching& \WinningStatesSizeThirtysecFruitCatching  \\
            HackAttack      & \MeanCovWaitTenHackAttack&  \WinningStatesWaitTenHackAttack        & \MeanCovWaitHundredHackAttack &\WinningStatesWaitHundredHackAttack        & \MeanCovWaitInfHackAttack& \WinningStatesWaitInfHackAttack          & \MeanCovOneMinHackAttack &\WinningStatesSizeOneminHackAttack          & \MeanCovThirtySecHackAttack &\WinningStatesSizeThirtysecHackAttack        \\
            Pong            & \MeanCovWaitTenPong & \WinningStatesWaitTenPong                    & \MeanCovWaitHundredPong &\WinningStatesWaitHundredPong                    & \MeanCovWaitInfPong& \WinningStatesWaitInfPong                      & \MeanCovOneMinPong &\WinningStatesSizeOneminPong                      & \MeanCovThirtySecPong &\WinningStatesSizeThirtysecPong                    \\
            Snake           & \MeanCovWaitTenSnake&  \WinningStatesWaitTenSnake                  & \MeanCovWaitHundredSnake &\WinningStatesWaitHundredSnake                  & \MeanCovWaitInfSnake &\WinningStatesWaitInfSnake                    & \MeanCovOneMinSnake& \WinningStatesSizeOneminSnake                    & \MeanCovThirtySecSnake& \WinningStatesSizeThirtysecSnake                  \\
            SpaceOdyssey    & \MeanCovWaitTenSpaceOdyssey & \WinningStatesWaitTenSpaceOdyssey    & \MeanCovWaitHundredSpaceOdyssey& \WinningStatesWaitHundredSpaceOdyssey    & \MeanCovWaitInfSpaceOdyssey& \WinningStatesWaitInfSpaceOdyssey      & \MeanCovOneMinSpaceOdyssey &\WinningStatesSizeOneminSpaceOdyssey      & \MeanCovThirtySecSpaceOdyssey &\WinningStatesSizeThirtysecSpaceOdyssey    \\
            \midrule
            Mean            & \TotalMeanCovWaitTen &\WinningStatesAverageWaitTen                & \TotalMeanCovWaitHundred & \WinningStatesAverageWaitHundred                & \TotalMeanCovWaitInf & \WinningStatesAverageWaitInf                  & \TotalMeanCovOneMin & \WinningStatesAverageSizeOnemin                  & \TotalMeanCovThirtySec & \WinningStatesAverageSizeThirtysec                \\
            \bottomrule
        \end{tabular}
    }
            \vspace{-1em}
\end{table}

Since the first experiment revealed that no-operation actions do not benefit the optimisation  process, we continue with a no-operation threshold of $\delta_t=\infty$ and analyse the effects of different training dataset sizes next.
\cref{fig:RQ1-Sizes} and \cref{tab:RQ1} indicate that a short recording length of only one minute achieves the fastest progress over the entire search duration and reaches the most winning states (\WinningStatesAverageSizeOnemin/30) with an average coverage of \TotalMeanCovOneMin\%.
These results deviate from other supervised learning scenarios where bigger datasets with more training samples usually improve the generalisation error, leading to overall better results. %\todo{Doesn't seem to be a justification for anything, do yu mean ``where'' maybe instead of ``since''?}
One reason for these results is related to our application scenario because large datasets  may decelerate the search progress by requiring increasingly more time for gradient descent.
As \cref{tab:Records} shows, recording for three minutes involves about three times more training samples than recording for one minute.
The increased dataset size may lead to more than a three-fold increase in gradient descent training durations since early stopping tends to interrupt the search far later due to higher generalisation capabilities. %\todo{Confusing sentence, it makes it difficult to comprehend whether that is a good or a bad thing.}
Because a higher degree of generalisation is usually beneficial, it becomes clear that human gameplay can only approximate our goal of evolving robust test input generators and overfitting on these traces should be avoided.
On the other hand, \cref{fig:RQ1-Sizes} indicates that recording for only 30 seconds generates too few training examples, which impairs the positive effect of applying gradient descent in the early stages of the search.

Overall, the experiments reveal that when adapting input generating neural networks to human gameplay traces, the search benefits the most from datasets excluding no-operations and of limited size, which deviates from other supervised learning scenarios.

\vspace{1em}

\noindent\fcolorbox{black}{black!5}{\parbox{.98\columnwidth}{\textbf{RQ1 (Recordings)}: Optimal gameplay traces for adapting input generating networks to human gameplay are based on a recording duration of 1 minute and do not contain no-operation actions.}}

\subsection{\mbox{RQ2 (NE+SGD): Does neuroevolution benefit} from gradient descent-based optimisation?}

\begin{figure*}[!tbp]
\centering
    \includegraphics[width=0.24\linewidth]{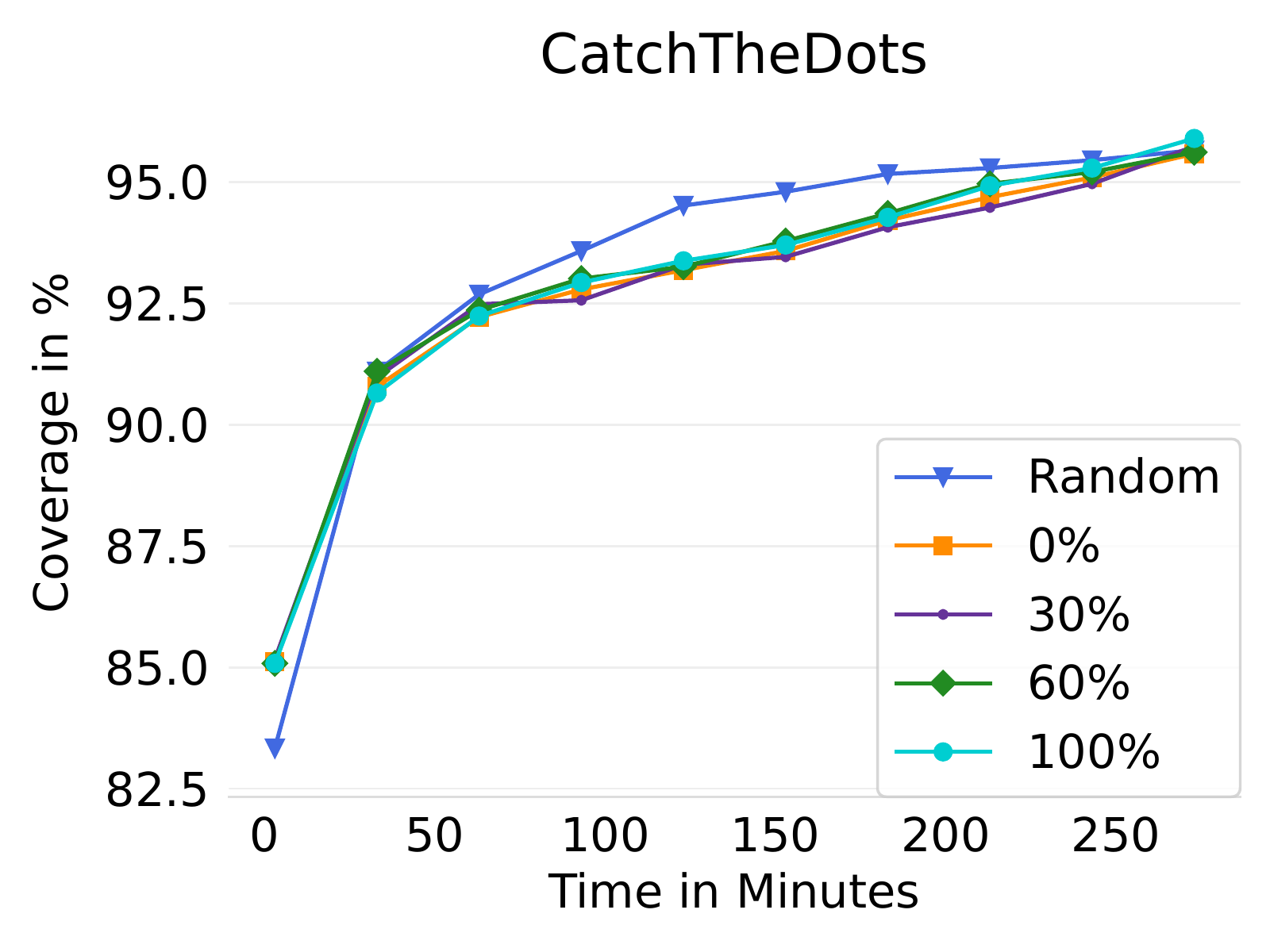}
    \includegraphics[width=0.24\linewidth]{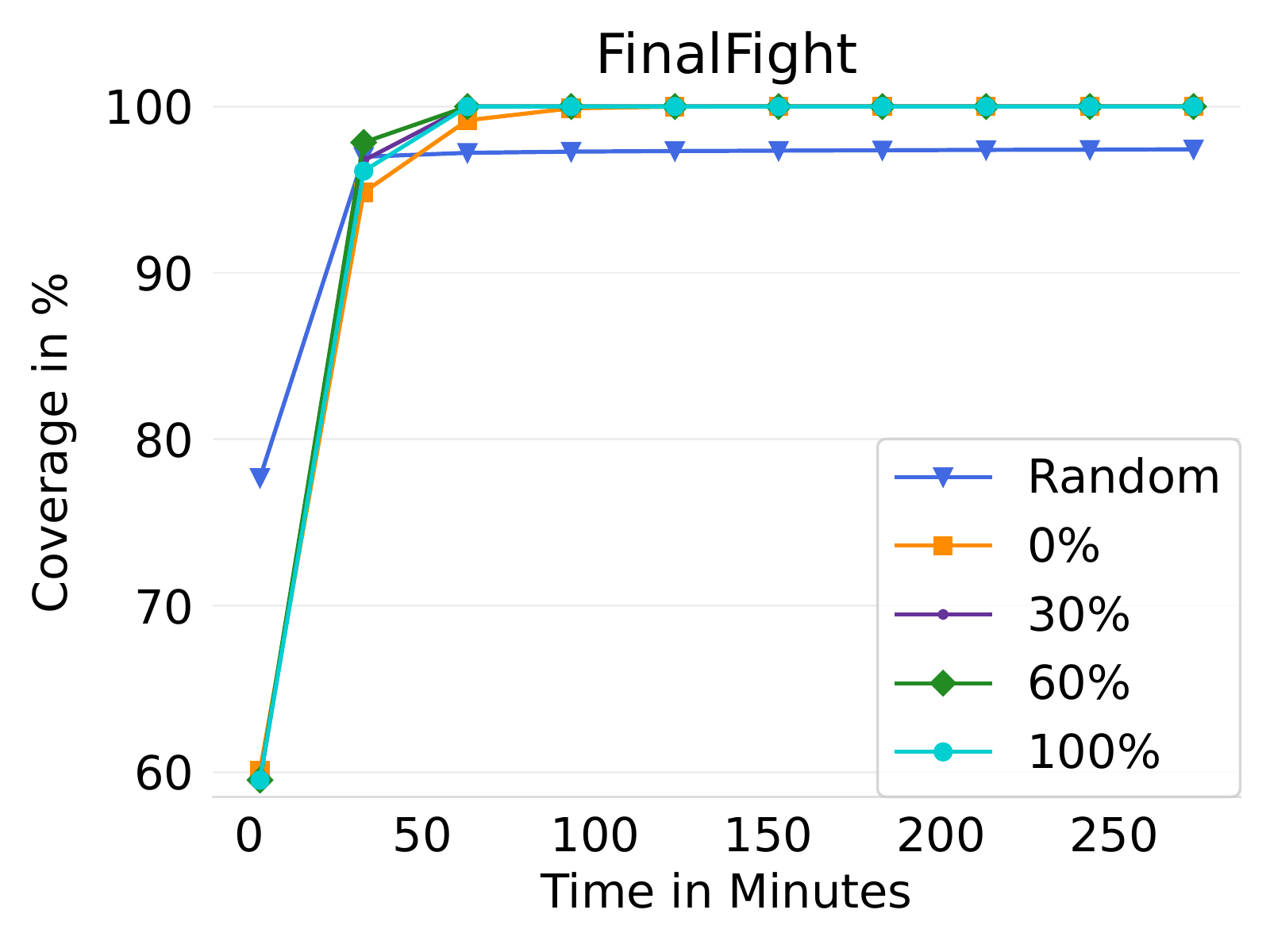}
    \includegraphics[width=0.24\linewidth]{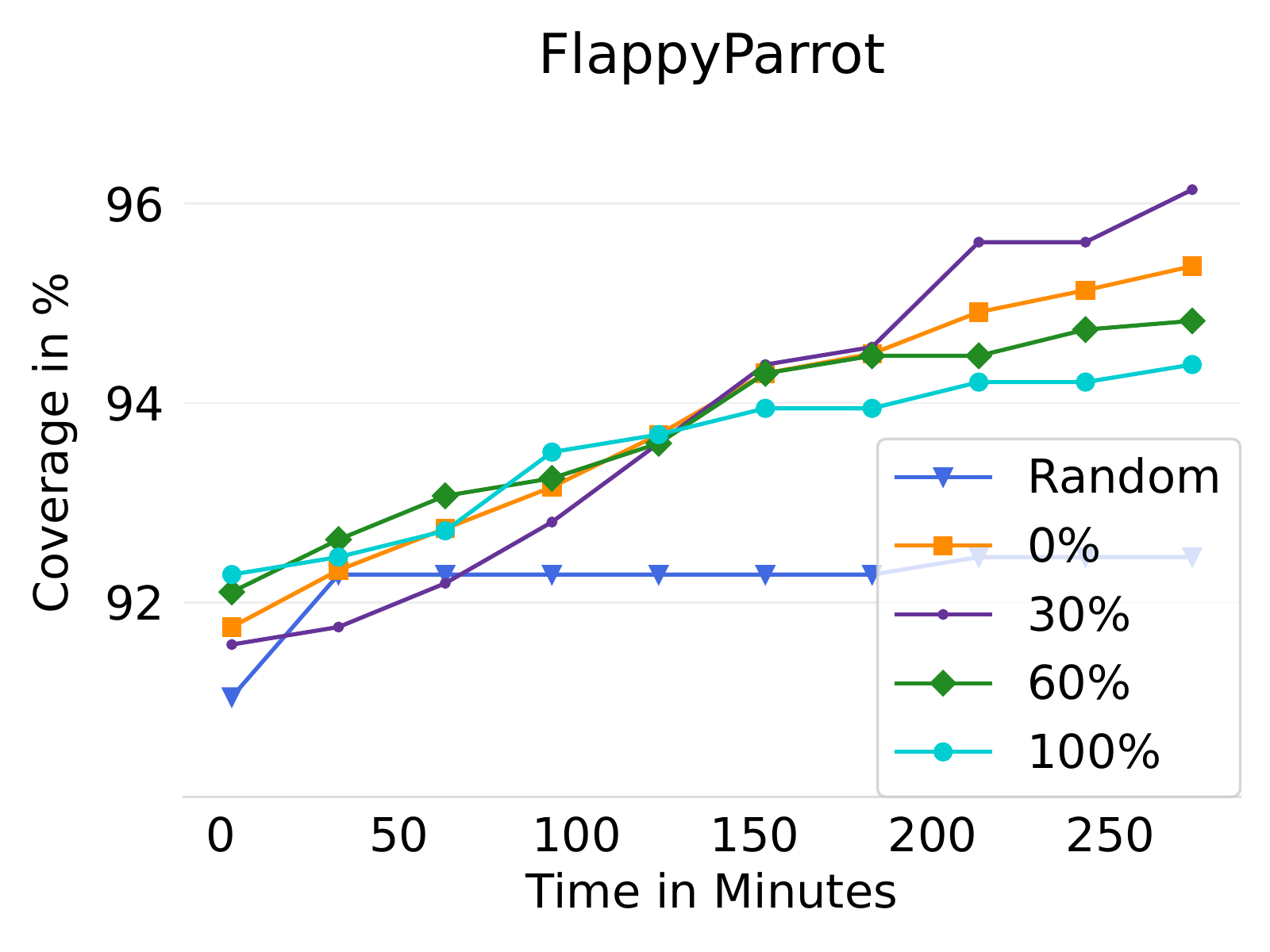}
    \includegraphics[width=0.24\linewidth]{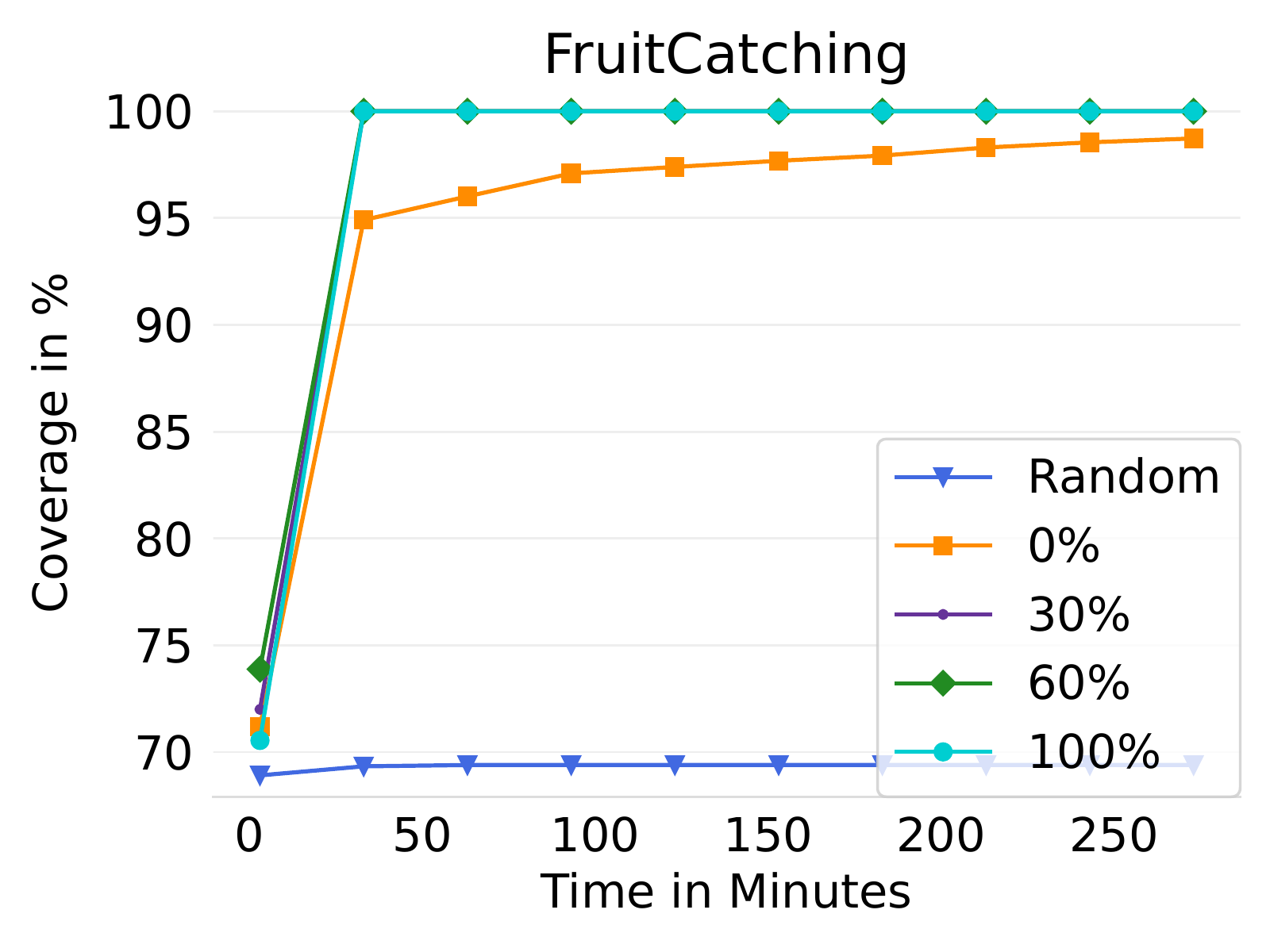}\par\medskip
    \includegraphics[width=0.24\linewidth]{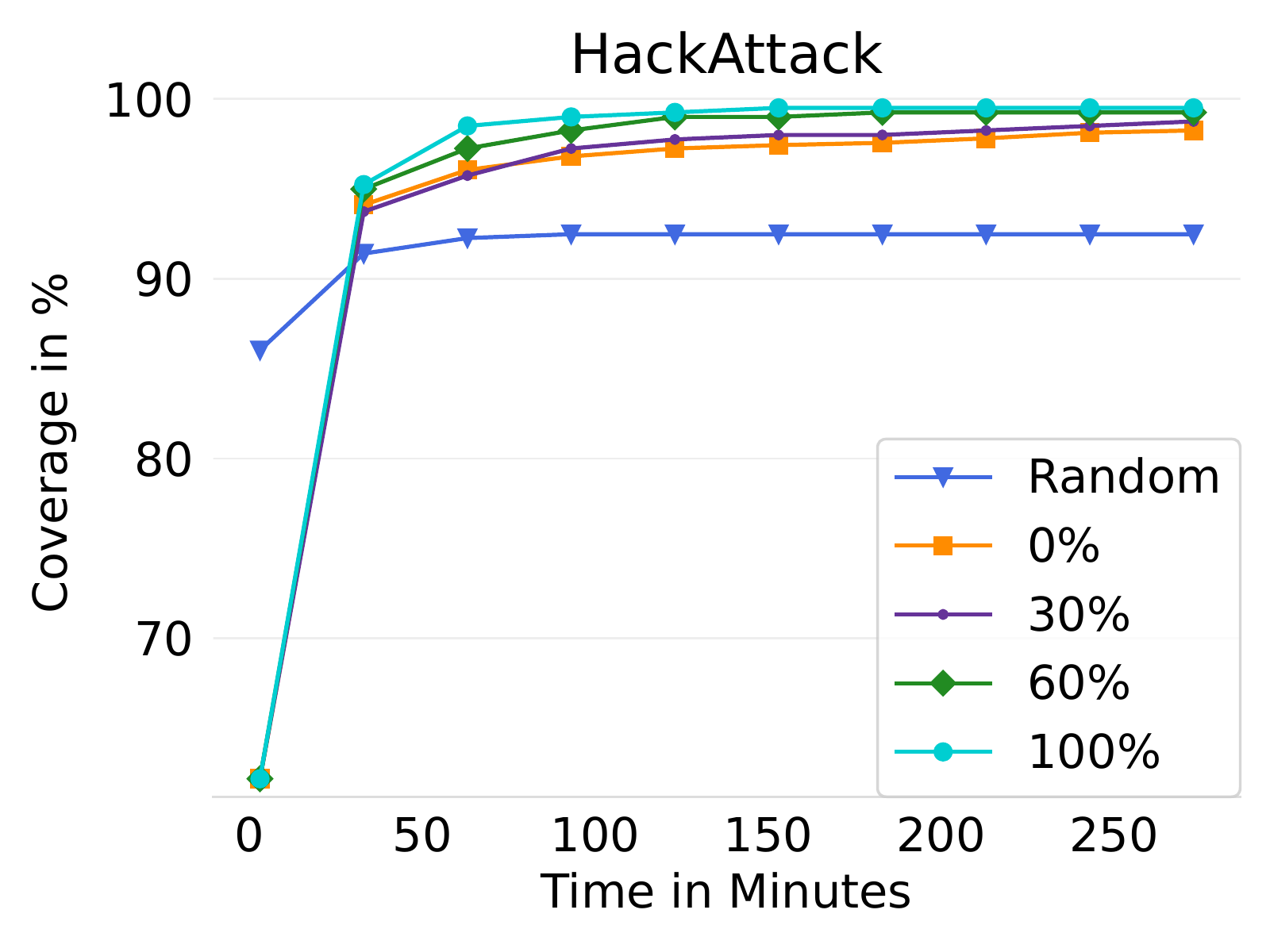}
    \includegraphics[width=0.24\linewidth]{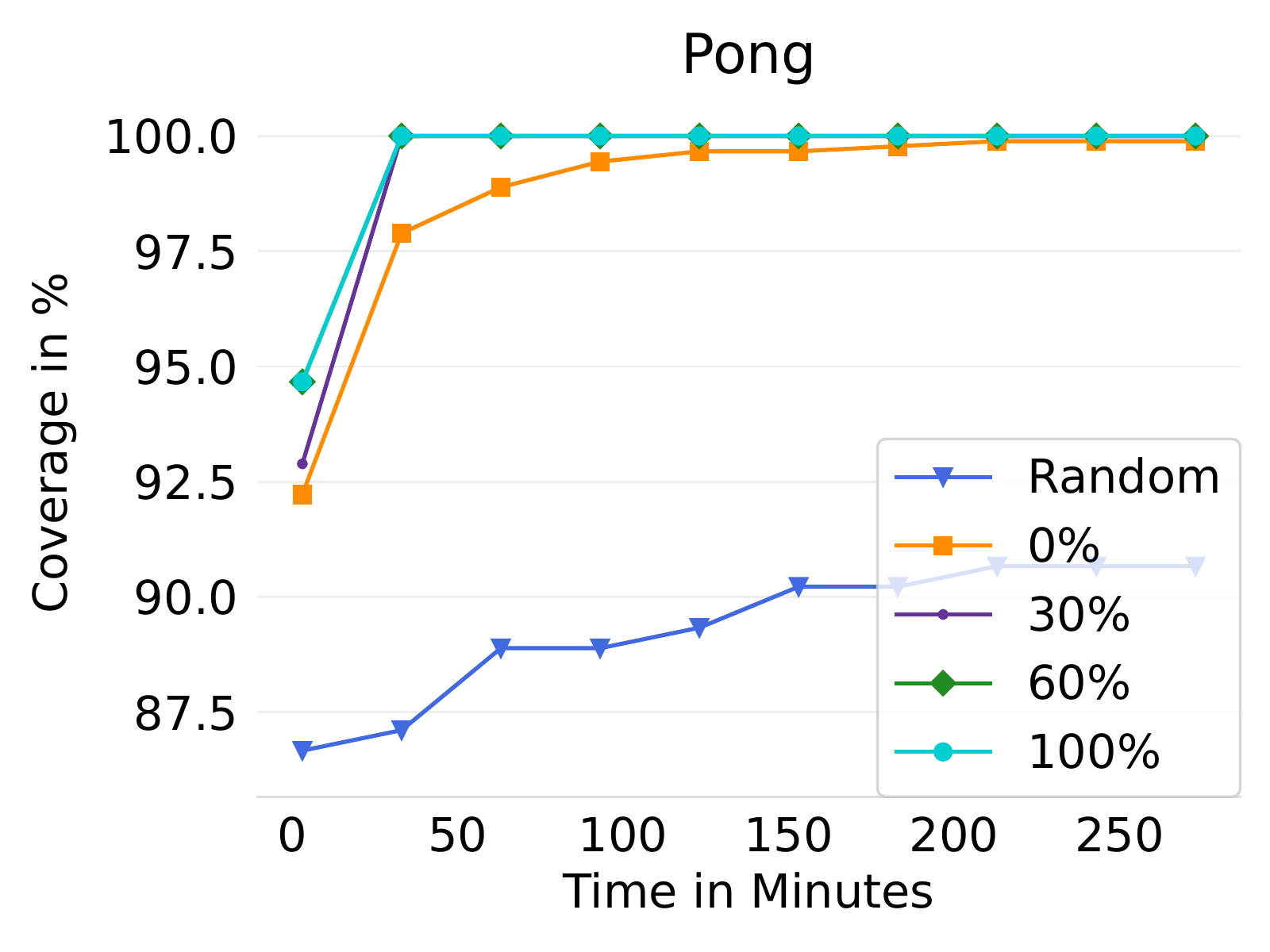}
    \includegraphics[width=0.24\linewidth]{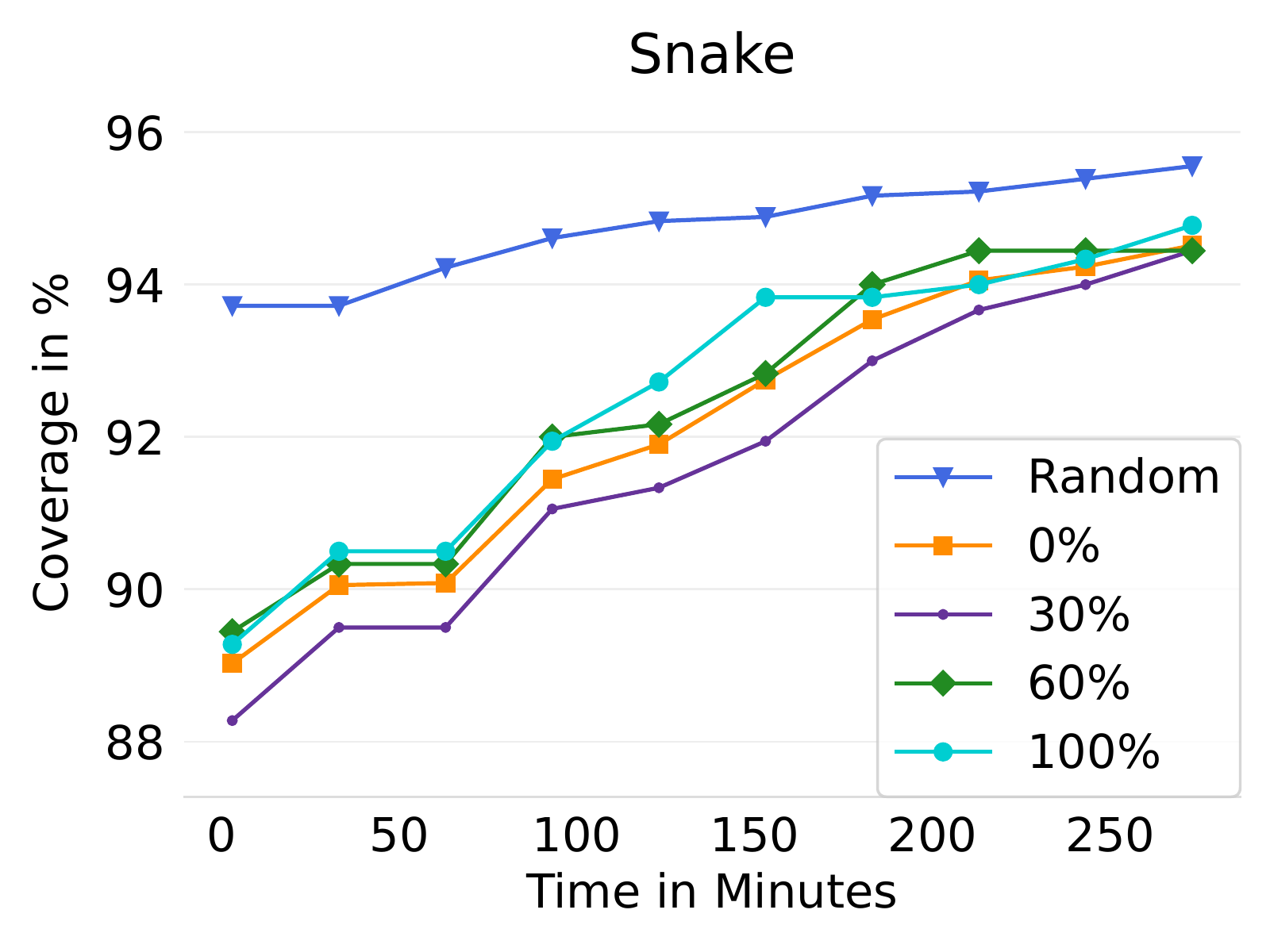}
    \includegraphics[width=0.24\linewidth]{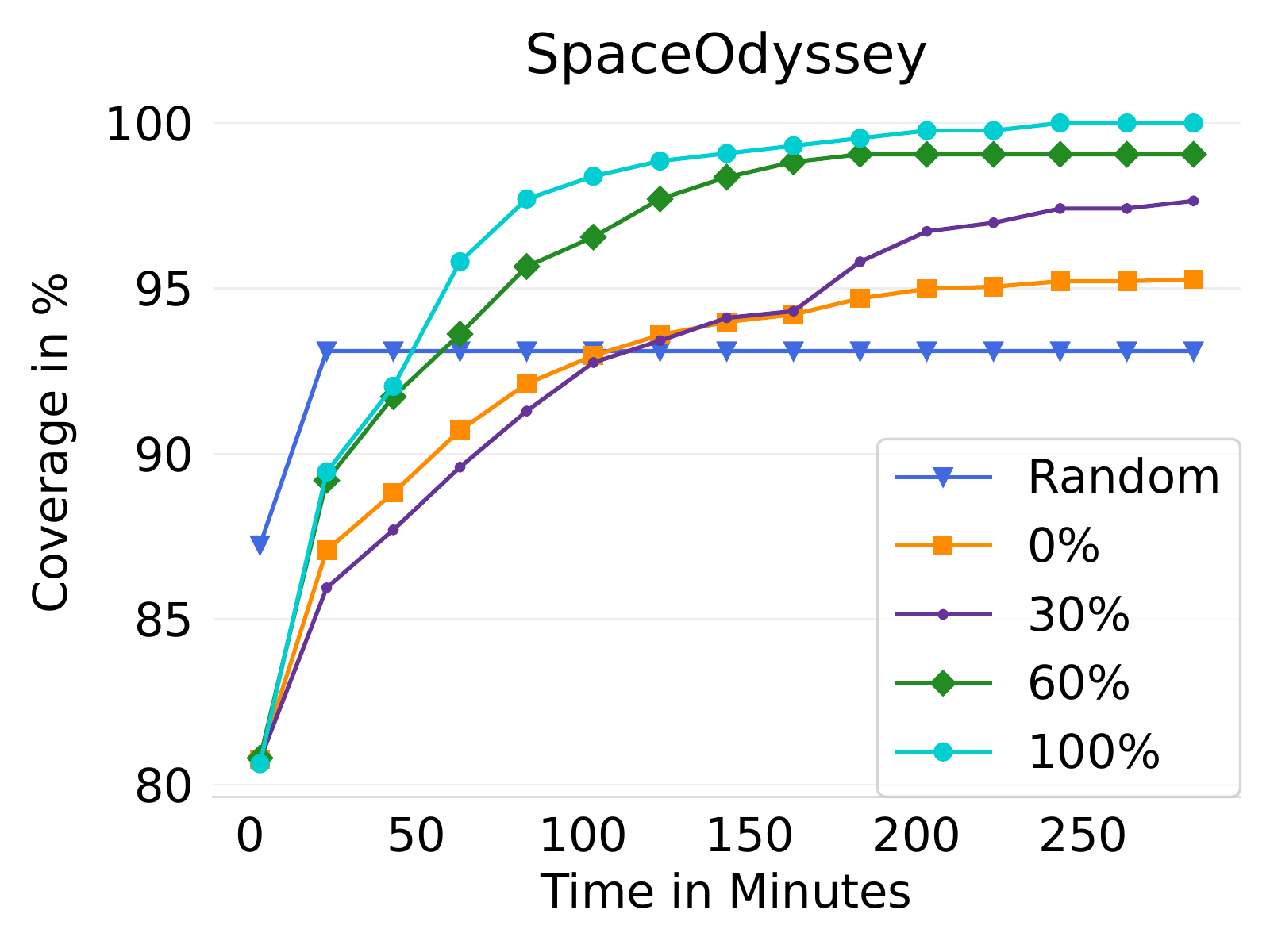}
\caption{\mbox{Coverage over time of the random test generator and \Neatest with varying probabilities of applying gradient descent.}}
\label{fig:RQ2}
\end{figure*}
RQ2 examines different probabilities of adapting test-input generating networks to human gameplay traces instead of applying the conventional weight mutation operation.
\Cref{fig:RQ2} shows that in most games, \Neatest achieves more coverage than the random tester in  all phases of the search, regardless of the gradient descent probability used.
However, in the games \emph{CatchTheDots} and \emph{Snake}, we can observe quite different results, with \Neatest reaching fewer statements than the random tester in the early stages of the search.
Due to a search duration of only five hours, some trivial statements like key-press handlers are not targeted during \Neatest's target selection process, and networks are not optimised to cover them.
Thus, they may not learn to cover these statements reliably, for instance, by pressing the same buttons regardless of the encountered program scenarios.
On the other hand, since the random tester generates static sequences that do not adapt to the game behaviour, it does not need to be optimised in the same way as networks to cover these simple statements.
However, \cref{tab:RQ2} shows that by optimising toward reaching challenging statements, the 100\% configuration, which discards the probabilistic weight mutation altogether, reaches these states in \emph{CatchTheDots} and \emph{Snake} more often (\WinningStatesHundredProbCatchTheDots and \WinningStatesHundredProbSnake times) than the random tester (\WinningStatesRandomCatchTheDots and \WinningStatesRandomSnake times). 

Overall, gradient descent is a valuable addition to neuroevolution, achieving an average coverage of \TotalMeanCovProbThirty\% (30\%), \TotalMeanCovProbSixty\% (60\%) and \TotalMeanCovProbHundred\% (100\%) across all programs and repetitions compared to the zero probability baseline (i.e., \Neatest without gradient descent) with \TotalMeanCovProbZero\%.
When interpreting coverage, keep in mind that the domain-specific nature of \Scratch blocks makes these values difficult to compare to regular programs: High coverage values can frequently be achieved without actual gameplay, and successful gameplay is often reflected by only small increases in coverage.

%\Cref{fig:RQ2} shows that high gradient descent probabilities of 60\% and 100\% achieve considerably more coverage in a shorter duration than the evolutionary search on its own in 6/8 \Scratch games.
Although there is little room for improvement, the 100\% configuration (i.e., only using gradient descent for mutation) reaches significantly more coverage than the 0\% baseline in \emph{HackAttack}, \emph{Snake} and \emph{SpaceOdyssey}.
Moreover, \cref{fig:RQ2} shows that gradient descent considerably reduces the required search time for   \emph{FruitCatching}, \emph{HackAttack}, \emph{Pong} and \emph{SpaceOdyssey}. %\todo{Maybe back this with numbers}
These speed-ups are possible because in these games the surrogate of gameplay traces aligns perfectly with the optimisation goal of generating robust input generators. 
For instance, precise mouse movements as required for \emph{Pong} (\cref{fig:Pong}) can be captured meticulously, allowing gradient descent to train the networks to perform precise mouse movements on the screen.

\begin{table}[t!]
    \caption{Average coverage (C) and reached winning states (W) with different gradient descent probabilities. Boldface indicates statistical significance with $p$ <  0.1 over the 0\% baseline.}
    \label{tab:RQ2}
    \vspace{-1em}
    \resizebox{\columnwidth}{!}{
        \setlength{\tabcolsep}{2pt} 
        \begin{tabular}{lrr@{\extracolsep{5pt}}rr@{\extracolsep{5pt}}rr@{\extracolsep{5pt}}rr@{\extracolsep{5pt}}rr}
            \toprule
            \multicolumn{1}{r}{}        & \multicolumn{2}{c}{Random}                       & \multicolumn{2}{c}{0\%}                                                            & \multicolumn{2}{c}{30\%}                                                 & \multicolumn{2}{c}{60\%}                                                  & \multicolumn{2}{c}{100\%}                                                   \\
                                                  \cline{2-3}                                       \cline{4-5}                                                                     \cline{6-7}                                                                     \cline{8-9}                                                                 \cline{10-11}
            Program         &                       C  &  W                                       &   C &  W                                                                    &              C          &  W                                              &        C   &  W                                                         &  C          &  W                                                                \\
            \midrule
            CatchTheDots       & \MeanCovRandomCatchTheDots & \WinningStatesRandomCatchTheDots          & \MeanCovProbZeroCatchTheDots & \WinningStatesZeroProbCatchTheDots     & \MeanCovProbThirtyCatchTheDots & \WinningStatesThirtyProbCatchTheDots     &  \MeanCovProbSixtyCatchTheDots & \WinningStatesSixtyProbCatchTheDots      & \MeanCovProbHundredCatchTheDots & \WinningStatesHundredProbCatchTheDots   \\
            FinalFight         & \MeanCovRandomFinalFight & \WinningStatesRandomFinalFight              & \MeanCovProbZeroFinalFight & \WinningStatesZeroProbFinalFight         & \MeanCovProbThirtyFinalFight & \WinningStatesThirtyProbFinalFight         &  \MeanCovProbSixtyFinalFight & \WinningStatesSixtyProbFinalFight          & \MeanCovProbHundredFinalFight & \WinningStatesHundredProbFinalFight       \\
            FlappyParrot       & \MeanCovRandomFlappyParrot & \WinningStatesRandomFlappyParrot          & \MeanCovProbZeroFlappyParrot & \WinningStatesZeroProbFlappyParrot     & \MeanCovProbThirtyFlappyParrot & \WinningStatesThirtyProbFlappyParrot     &  \MeanCovProbSixtyFlappyParrot & \WinningStatesSixtyProbFlappyParrot      & \MeanCovProbHundredFlappyParrot & \WinningStatesHundredProbFlappyParrot   \\
            FruitCatching      & \MeanCovRandomFruitCatching & \WinningStatesRandomFruitCatching        & \MeanCovProbZeroFruitCatching & \WinningStatesZeroProbFruitCatching   & \MeanCovProbThirtyFruitCatching & \WinningStatesThirtyProbFruitCatching   &  \MeanCovProbSixtyFruitCatching & \WinningStatesSixtyProbFruitCatching    & \MeanCovProbHundredFruitCatching & \WinningStatesHundredProbFruitCatching \\
            HackAttack         & \MeanCovRandomHackAttack & \WinningStatesRandomHackAttack              & \MeanCovProbZeroHackAttack & \WinningStatesZeroProbHackAttack         & \MeanCovProbThirtyHackAttack & \WinningStatesThirtyProbHackAttack         &  \MeanCovProbSixtyHackAttack & \WinningStatesSixtyProbHackAttack          & \MeanCovProbHundredHackAttack & \WinningStatesHundredProbHackAttack       \\
            Pong               & \MeanCovRandomPong & \WinningStatesRandomPong                          & \MeanCovProbZeroPong & \WinningStatesZeroProbPong                     & \MeanCovProbThirtyPong & \WinningStatesThirtyProbPong                     &  \MeanCovProbSixtyPong & \WinningStatesSixtyProbPong                      & \MeanCovProbHundredPong & \WinningStatesHundredProbPong                   \\
            Snake              & \MeanCovRandomSnake & \WinningStatesRandomSnake                        & \MeanCovProbZeroSnake & \WinningStatesZeroProbSnake                   & \MeanCovProbThirtySnake & \WinningStatesThirtyProbSnake                   &  \MeanCovProbSixtySnake & \WinningStatesSixtyProbSnake                    & \MeanCovProbHundredSnake & \WinningStatesHundredProbSnake                 \\
            SpaceOdyssey       & \MeanCovRandomSpaceOdyssey & \WinningStatesRandomSpaceOdyssey          & \MeanCovProbZeroSpaceOdyssey & \WinningStatesZeroProbSpaceOdyssey     & \MeanCovProbThirtySpaceOdyssey & \WinningStatesThirtyProbSpaceOdyssey     &  \MeanCovProbSixtySpaceOdyssey & \WinningStatesSixtyProbSpaceOdyssey      & \MeanCovProbHundredSpaceOdyssey & \WinningStatesHundredProbSpaceOdyssey   \\
            \midrule
            Mean               & \TotalMeanCovRandom & \WinningStatesAverageRandom                      & \TotalMeanCovProbZero & \WinningStatesAverageZeroProb                 & \TotalMeanCovProbThirty & \WinningStatesAverageThirtyProb                 &  \TotalMeanCovProbSixty & \WinningStatesAverageSixtyProb                  & \TotalMeanCovProbHundred & \WinningStatesAverageHundredProb               \\
            \bottomrule
        \end{tabular}
    }
    
\end{table}

If the surrogate does not align well with the input generation goal, this may have negative consequences, as can be seen for \emph{FlappyParrot} depicted in~\cref{fig:Dataset}.
In this game, players have to manoeuvre a parrot through pipes by pressing the space bar to command the bird to fly higher.
Since the game traces consist of only one viable action, gradient descent teaches the networks to continuously perform the same action instead of teaching the networks to execute the respective action only at the right moments.
High probabilities of applying gradient descent may lead to less diverse populations as most networks are trained to align with the given gameplay record.

By adapting the networks to human gameplay traces, we achieve the same amount of coverage (\TotalMeanCovProbZero\%) as \Neatest without gradient descent reaches in five hours already after \SpeedOverBaselineProbThirty, \SpeedOverBaselineProbSixty\ and \SpeedOverBaselineProbHundred\ minutes when applying a gradient descent probability of 30\%, 60\% and 100\%, respectively.
Nevertheless, completely replacing the probabilistic weight mutation with the systematic gradient descent operation might cause a lack of population diversity and a decrease in search performance if the gameplay traces do not align well with covering a specific program statement, such as in the \emph{FlappyParrot} game. %\todo{What in the data suggests this?}
 \vspace{1em}

 \noindent\fcolorbox{black}{black!5}{\parbox{.95\columnwidth}{\textbf{RQ2 (NE+SGD)}: Combining neuroevolution with gradient descent reduces the required search time by up to 83\% but should be used with caution due to lower population diversity.}}

\section{Related Work}
Gradient descent has previously been combined with neuroevolution for optimising network weights in conjunction with the architecture of a neural network.
For instance, \emph{NEAT+Q}~\cite{whiteson2006evolutionary} and \emph{Online NEAT+Q}~\cite{whiteson2006online} propose evolutionary function approximation by optimising neural networks toward approximating value functions used by \emph{Q-Learning}~\cite{watkins1989learning}.
\emph{Learning-NEAT (L-NEAT)}~\cite{chen2006neuroevolution} combines gradient descent with neuroevolution to solve the two classification tasks of iris flower prediction~\cite{fisher1936use} and scale balancing. 
Finally, Desell~\cite{desell2017large} proposes the \emph{Evolutionary eXploration of Augmenting Convolutional Topologies (EXACT)} method for optimising neural networks toward recognising handwritten digits in the MNIST~\cite{lecun1998gradient} dataset. 
These approaches differ fundamentally from our proposed approach as they use training datasets that closely resemble the given problem domain, which enables an improved weight optimisation process by replacing the weight mutation operator entirely with gradient descent.
When testing games, such a fitting dataset is infeasible, and we have to approximate a training dataset via gameplay traces.
Thus, we refrain from discarding the weight mutation operator and instead integrate gradient descent as an alternative to the evolutionary search for network weights.

Neuroevolution has been used to tackle many problems that are related to video games~\cite{risi2015neuroevolution}.
For instance, it has been applied to predict which level in a game a player prefers to play next~\cite{pedersen2010modeling} or even to synthesise personalised levels altogether~\cite{shaker2010towards}.
Other use cases include automatic content generation~\cite{hastings2009automatic}, simulating human gameplay~\cite{schrum2011ut, schrum2013human} or constructing complex NPC behaviour via multi-objective neuroevolution~\cite{schrum2008constructing}.
However, the most widespread application scenario is to use neuroevolution in order to train networks toward mastering games~\cite{hausknecht2014neuroevolution, butz2009optimized, schrum2018evolving}.
In such approaches, neural networks are evolved to either serve as state-action evaluators~\cite{fogel2004self} that evaluate the future state of an action and choose the action that promises the best outcome or as direct-action selectors~\cite{togelius2009super, togelius2005evolving} by deriving actions directly from the current program state.
 Many strategies have been proposed for mastering video games, such as incremental evolution~\cite{togelius2006evolving}, transfer learning~\cite{cardamone2011transfer} and competitive coevolution~\cite{lucas2005evolving}.
 However, to the best of the author's knowledge, neuroevolution has never been combined with gradient descent-based weight optimisation in order to improve the learning of general gameplaying or game testing.

 The integration of domain knowledge into evolutionary search is
 commonly referred to as \emph{seeding} and has been used in search-based testing~\cite{rojas2016seeding}, for example
 by integrating literal values that the search would otherwise
 struggle to generate. While this is similar in spirit to our
 approach, we do not integrate raw values but rather use traces to
 inform weight updates. This is akin to memetic algorithms,
 which have been used for test data generation by combining
 global and local search~\cite{harman2009theoretical,fraser2015memetic} or global
 search and dynamic symbolic execution~\cite{malburg2014search,galeotti2013improving}.
 Our approach differs by combining global search with gradient descent based weight updates.

%\subsection{Testing Games}

%Nowadays, virtually all video games are tested manually or semi-automatically~\cite{politowski2021survey}.
%Furthermore, most manual testing approaches do not even focus on systematically discovering bugs but on evaluating the usability of games~\cite{desurvire2004using}.
%For example, Diah et al.~\cite{diah2010usability} propose an observation method for preschool children aged between five and six years, Korhonen et al.~\cite{korhonen2006playability} evaluate the usability of mobile games, and Ferre et al.~\cite{ferre2009playability} introduce heuristics to assess the playability of a sport game.
%On the other hand, semi-automated game testing frameworks such as the scenario-based Venus II~\cite{cho2010online}, the beta tester Crushinator~\cite{schaefer2013crushinator}, and a test framework for early game prototypes~\cite{smith2009computational} focus on automating the test execution process but developers still have to generate test cases themselves.
%There are a few approaches capable of generating test input sequences automatically like the online combat game tester Wuji~\cite{zheng2019wuji} and the behaviour testing framework of Chan et al.~\cite{chan2004evolutionary}.
%However, these automated approaches focus on specific game genres and lack the amount of generisability offered by Neatest~\cite{feldmeier2022neuroevolution}.
%Furthermore, none of these testing frameworks covers program statements systematically using fitness functions derived from the research field of search-based software testing.
\section{Conclusions}

Generating adequate test inputs for games is challenging because many program statements in games tend to be hard to reach and because games are heavily randomised.
\Neatest solves these challenges by generating test suites of neural networks that learn to play games meaningfully while adapting to changes in program behaviour.
In this paper, we demonstrated the use of systematically recorded human gameplay sessions as a data source, used in an extension of the \Neatest approach that combines neuroevolution with gradient descent-based weight optimisation. Experiments on eight \Scratch games suggest that combining neuroevolution with gradient descent-based weight optimisation improves the search progress by reaching more coverage in less time.

With this work, we laid a first foundation for combining
neuroevolution with gradient-descent to
evolve networks that serve as robust test input generators for games.
In future work, we envision the evaluation of several
regularisation and learning rate adaptation techniques that may
improve the benefit of including gradient descent in the evolutionary
search even more. Furthermore, the design space of integrating
gradient descent into \Neatest allows for further variations that
could be explored, such as changing the order in which
statements are explored based on the trace contents.
Finally, we envision the combined use of neuroevolution and gradient descent in other dynamic domains such as automated driving, robot control and testing of control systems.

\begin{acks}
This work is supported by \grantsponsor{FR 2955/3-1}{DFG project FR2955/3-1 “TENDER-BLOCK: Testing, Debugging, and Repairing Block-based Programs”}{https://gepris.dfg.de/gepris/projekt/418126274}. The authors are responsible for this publication's content.
\end{acks}

\balance

% Bibliography
\bibliographystyle{ACM-Reference-Format}
  \bibliography{main}
\end{document}